\documentclass[floatfix,prb,twocolumn,showpacs,preprintnumbers,amsmath,amssymb]{revtex4-2}
\usepackage{epsfig}
\usepackage{color}
\usepackage{amssymb}
\usepackage{amsmath}
\usepackage[ngerman,english]{babel}
\usepackage{multirow}
\usepackage{graphicx}
\usepackage{subfigure}
\usepackage{ulem}

\newcommand{\be}{\begin{equation}}
\newcommand{\ee}{\end{equation}}
\newcommand{\bea}{\begin{eqnarray}}
\newcommand{\eea}{\end{eqnarray}}

\newcommand{\GR}[1]{\textcolor{black}{#1}}

\begin{document}

\title{The virial expansion of plasma properties: benchmarks for numerical results}
\author{G. R\"opke \footnote{Orcid {0000-0001-9319-5959}}}
\affiliation{Institut f\"ur Physik, Universit\"at Rostock, D-18051 Rostock, Germany}%\\
 \email{gerd.roepke@uni-rostock.de}

\date{\today}
\begin{abstract}
Expressions for the thermodynamic and transport properties of plasmas are derived from quantum statistics in the form of equilibrium correlation functions. These can be evaluated using analytical methods or numerical approaches such as DFT-MD or PIMC simulations. Virial expansions are obtained using the Green's function method. They provide benchmarks for numerical simulations and are useful in the low-density range. The results for the equation of state are discussed for the uniform electron gas and the hydrogen plasma.
Transport properties such as the dielectric function are also of interest.
Virial expansions are considered for the electrical direct current conductivity as a special case of the dielectric function. Examples are given and it is explained where further work is needed to obtain a consistent description of the properties of hot and dense plasmas.
\end{abstract}

\maketitle

\section{Introduction}\label{sec0}

Plasma properties are of interest for technical applications and for describing astrophysical objects.
We focus on the simplest case, the uniform electron gas (UEG) and the hydrogen plasma, see \cite{WDM,Bonitz24}.
The equilibrium state is determined by the temperature $T$ and the electron number density $n$.
The UEG consists of electrons with the charge $-e$ moving in a homogeneous background with a charge density of $e n$ to ensure the charge neutrality of the system. 
The H plasma consists of electrons and protons having the same density $n_e=n_p=n$.
The Hamilton operator contains the kinetic energy of the electrons and, in the case of H plasma, that of the protons.
The interaction is given by the Coulomb potential
\begin{equation}
V(r_i,r_j) = \frac{1}{2} \sum_{i\neq j} \frac{e_ie_j}{4 \pi \epsilon_0 \,|{\bf r}_i-{\bf r}_j|}
\end{equation}
between particles with charge $e_i,e_j$ at position ${\bf r}_i,{\bf r}_j$. 
For the UEG, where we only have electrons, the interaction with the homogeneously charged background must be added.
Charge neutrality is necessary to avoid divergences in the thermodynamic limit, where a homogeneous system is considered and the volume approaches infinity.
Generalisations to more complex plasmas such as other elements or mixtures are possible using the same methods described here.

The equilibrium state of these Coulomb systems is specified in the grand canonical ensemble, defined by $\beta=1/k_{\rm B}T$ and the chemical potentials $\mu_i$. 
The statistical operator is
\begin{equation}
\label{rhoeq}
\rho_{\rm eq}=e^{-\beta (\hat H-\sum_i\mu_i \hat N_i)}/{\rm Tr}\left[e^{-\beta (\hat H-\sum_i\mu_i \hat N_i)}\right].
\end{equation}
We use second quantisation with creation/annihilation operators $a^+_{k},a^{}_k$ in single-particle states $|k\rangle$,
where $k=\{ {\bf k}, \sigma,c \}$ describes not only the wave number $\bf k$ but also the spin $\sigma$ and the species $c=e, p$.
The Hamiltonian contains the kinetic energy $\sum_k E_k a_k^+a^{}_k$, $E_k =\hbar^2 k^2/(2 m_c)$, and the interaction
\begin{equation}
 \frac{1}{2} \sum_{kk' {\bf q}} \frac{e_ce_{c'}}{\epsilon_0\,\Omega_0\,\, q^2}a_{k+q}^+a_{k'-q}^+a_{k'}^{}a_k^{}.
\end{equation}
The wave number vectors $\bf k,k'$  are changed by the transition wave vector $\bf q$, spin and species remain unchanged.
$\Omega_0$ denotes the volume of the system ($V$ was already taken for the potential energy), we have $\sum_{\bf k}=\int d^3 k \,\,{\Omega_0}/(2 \pi)^3$.
The particle number operator is $\hat N_i=\sum_k a_k^+a^{}_k \delta_{i,k}$ with given $i=\{\sigma, c\}$.
We prefer this detailed description; often some terms are omitted.

The physical properties of the plasma are expressed by mean values $\langle \hat B \rangle={\rm Tr}[\rho \hat B]$ of the corresponding operators $\hat B$.
In second quantization, these operators $\hat B$ are represented by the operators $a^+_{k},a^{}_k$. 
This means that we have to calculate mean values of products of creation and annihilation operators, the so-called correlation functions.
We therefore have to deal with two problems:
\begin{enumerate}
\item Which correlation functions must be used to calculate a physical property?
\item What methods can be used to calculate the correlation functions?
\end{enumerate}

Problem (1) is a physics question. For thermodynamics, we must consider the conserved quantities energy $\hat H$ and particle numbers $\hat N_i$ given above. 
The mean values $U(T,\mu_i)=\langle \hat H \rangle$ and $N_i(T,\mu_i)=\langle \hat N_i \rangle$ are referred to as equations of state (EoS). 
We consider these thermodynamic properties in Section \ref{sec:thermodynamics}.
For transport properties, we must take into account fluctuations in equilibrium according to linear response theory. 
For example, current-current fluctuations are expressed by $\langle \hat n_k(t) \hat n_{k'} \rangle$, where $\hat n_k= a_k^+a^{}_k$ is the occupation number of the state $|k\rangle$, and the $t$-dependence is given by the Heisenberg picture.
We consider transport properties in Section \ref{sec:transport}.

Problem (2) is a mathematical question. Since this is a many-particle problem, closed solutions are not possible. 
We can use the perturbation theory of quantum statistics, which is briefly described in section \ref{sec:green}.
Analytical expressions are obtained, which in general are approximations. 
Exact results can be found in limiting cases, for example the virial expansions.
An alternative is numerical simulations to solve the many-particle problem. 
Density-functional theory (DFT) for the electron system together with molecular dynamics (MD) for the ion systems  are frequently used.
A fundamental problem is the treatment of electron-electron correlations, which are taken into account in certain approximations, such as the density functional of the exchange-correlation energy.
This problem is circumvented by the path integral Monte Carlo (PIMC) approach, which solves quantum theory exactly (see Ref. \cite{B20} and references given therein).
However, due to the finite size of 
the particle numbers and the sign problem, the accuracy of PIMC simulations is limited.
We will not discuss these numerical methods in detail here. They are well described in review articles.

The Green's function method was developed in field theory and later successfully applied to quantum statistics.
There are very efficient approximations that describe physical effects, such as mean-field approximations, the quasi-particle concept, screening, the formation and dissociation of bound states, and Pauli blocking effects.
A description of this method can be found in monographs, see \cite{FW,Buch}, with specific application to Coulomb systems in \cite{KB,KKER86}.
We will only give a brief overview here in Sect. \ref{sec:green}.

Our aim is to use analytical results for comparison with numerical methods.
Virial expansions are benchmarks that can be used to verify the validity and accuracy of numerical results.
They are very efficient in limiting cases, for example for plasmas at low density.
It is also interesting to check the convergence range of the virial expansions.
Analytical results are essential for creating interpolation formulas that describe physical properties in reasonable  approximations.

\GR{Virial expansions are expansions in terms of density.
 How many terms of this expansion must be taken into account in order to describe the physical properties of warm and dense matter at finite density and temperature with a certain degree of accuracy?
The applicability of virial expansions was discussed for conductivity in \cite{cond25}.
The lowest-order terms  of the virial expansion provide acceptable results in the density-temperature range where free electrons are not degenerate and the formation of bound states can be neglected.
For thermodynamic properties, the applicability of virial expansions is limited by the formation of bound states, see \cite{RCER25}. 
As an alternative, the virial expansion of the quasi-particle energies of the free and bound components was proposed there.}

\GR{ This work is partly a review summarising similar results from previous publications,
but it also provides new, original results and points to future directions.
Our goal is to present this method of benchmarking numerical simulations for various physical quantities.
The treatment of H plasma, Sect. \ref{sec:TD-H},  is a brief presentation of the current work \cite{RCER25} and the references given therein. The discussion of the UEG, Sect. \ref{sec:UEG}, contains new results, for example regarding the convergence of the virial expansion.
The conductivity of H plasmas, Sect. \ref{sec:sigma},  is a brief review of various analyses, see \cite{pop24} and the references given therein.
The dielectric function, Sect. \ref{sec:df}, and related quantities such as the dynamic structure factor are more general properties that cover both thermodynamic quantities and conductivity. The treatment of the dielectric function is discussed, but results beyond the lowest order of the virial expansion of this complex quantity cannot be presented in this work.
The new results and directions for future research are summarised in the conclusions, Sect. \ref{sec:concl}.}

\section{Thermodynamic properties}
\label{sec:thermodynamics}

A fundamental quantity for describing the physical properties of a many-particle system is the spectral function, which is well defined even in a dense system.   The Fourier transform of the mean value of the creation of an electron at time $t$ in state $|k\rangle$ and the extraction of the electron at time $t'$ in state $|k'\rangle$ (Heisenberg picture)
\begin{equation}
\int d(t'-t) e^{i\omega (t'-t)} \langle a^+_{e,k}(t) a^{}_{e,k'}(t')\rangle =\frac{1}{e^{\beta \omega}+1}A_e(k,\omega ) \delta_{k,k'}
\end{equation}
defines the spectral function $A_e(k,\omega)$ of the electrons (the variables $T, \mu_i$ are not shown). Due to homogeneity in space and time (thermodynamic equilibrium), it is diagonal in the momentum representation $k$ and depends only on $(t'-t)$.  We can rewrite $n_e=\Omega_0^{-1} \sum_k\langle a^+_{e,k} a^{}_{e,k} \rangle$  for the electrons as 
\begin{eqnarray}
n_e(T, \mu_i) &=&\frac{1}{\Omega_0}\sum_k\int \frac{d \omega}{2 \pi} \frac{1}{e^{\beta \omega}+1}A_e(k,\omega ) \nonumber \\
&=&\int \frac{d \omega}{2 \pi}  \frac{1}{e^{\beta \omega}+1}D_e(\omega ), 
\end{eqnarray}
where 
\begin{equation}
    D_e(\omega )=\frac{1}{\Omega_0}\sum_k A_e(k,\omega) =\sum_\sigma\int \frac{d^3 k}{(2 \pi)^3} A_e(k,\omega)
\end{equation}
denotes the electron density of  states.
Similar expressions can be obtained for ions (protons), which, however, are considered to be classical particles under the conditions discussed in this work.

With the EoS $n_i(T,\mu_j)$, we can obtain other thermodynamic functions. 
 An inversion, i.e. solving this set of equations for the chemical potentials, gives us the relationships $\mu_i(T,n_j)$, such that temperature and density are the independent variables (canonical ensemble).
 In particular, the free energy density $f(T,n)$ which is a thermodynamic potential, is obtained 
by integration 
\begin{equation}
\label{fp(n)}
f(T,n)=\int_0^n \mu(T,n') dn' .%\,\,\, \quad\quad\quad p(T, \mu)=\int_{-\infty}^\mu n(T,\mu') d \mu'.
%f(T,n_i)=\int_0^{n_j} \mu_j(T,n'_j,n_i) dn'_j|_{n_i,i \neq j}, %\,\,\, \quad\quad\quad p(T, \mu)=\int_{-\infty}^\mu n(T,\mu') d \mu'.
\end{equation}
All  other thermodynamic variables are obtained from this quantity.

\subsection{Virial expansion and the generalized Beth-Uhlenbeck equation}
\label{sec:green}

The Green's function method is a perturbation theory approach. The convergence behavior of this expansion in powers of the interaction is not rigorously known.
The technique of Feynman diagrams makes it possible to describe many-particle effects by means of systematic perturbation expansion, including partial summations. 
Concepts such as quasiparticles, screening and the formation of bound states are introduced systematically.

\GR{ In the framework of thermodynamic Green's functions \cite{KB,FW,KKER86,Buch}, the single-particle Green function $G_i(k,iz_\lambda)$ is introduced, defined at the fermionic Matsubara frequencies $z_\lambda= \pi \lambda/\beta; \lambda =  \pm 1, \pm 3, \dots,$  
and $i=\{\sigma,c\}$ denotes spin and species.
$G_i(k,iz_\lambda)$ is related to the self-energy $\Sigma_i(k,iz_\lambda)$,
\begin{equation}
G_{i}(k,iz_\lambda)=\frac{1}{E_{i,k}+\Sigma_{i}(k,iz_\lambda)-\mu_{i}-iz_\lambda},%\quad E_{{i},k}=\frac{\hbar^2 k^2}{2 m_{i}}
\end{equation}
$E_{{i},k}=\hbar^2 k^2/2 m_{i} $ is the kinetic energy.
After analytical continuation in the complex $z$ plane, $iz_\lambda \to z$,} the spectral function  is obtained,
\begin{eqnarray}
A_i(k,\omega) &=& 2 \lim_{\epsilon \to 0} {\rm Im} G_i(k,\omega-i \epsilon)\\
=&&\!\!\!\!\!\!2 \frac{{\rm Im} \Sigma_i(k,\omega)}{[E_{i,k}+{\rm Re}\Sigma_i(k,\omega)-\mu_i-\omega]^2+[{\rm Im} \Sigma_i(k,\omega)]^2}.\nonumber
\end{eqnarray}
We focus on the electron spectral function and omit the index $i$. For small ${\rm Im}\, \Sigma(k,\omega)$, we obtain the expansion
\begin{eqnarray}
\label{Aexpans}
A(k,\omega) &\approx&\frac{2 \pi \delta(\omega+\mu_e-E_k^{\rm qu})}{1-\frac{d}{dz}{\rm Re}\Sigma(k,z)|_{z=E^{\rm qu}_k-\mu_e}} \nonumber \\&&+2{\rm Im} \Sigma(k,\omega+i0) \frac{d}{d \omega}\frac{{\cal P}}{\omega+\mu_e-E_k^{\rm qu}}.
\end{eqnarray}
The first term is the quasiparticle peak, a sharp $\delta$-like peak at the quasiparticle energy
\begin{equation}
\label{quEnGl}
E_k^{\rm qu}= E_k+{\rm Re}\Sigma(k,\omega)|_{\omega = E^{\rm qu}_k-\mu_e}.
\end{equation}
Correlations, in particular bound states, are obtained from the second part of 
equation (\ref{Aexpans}). 
For this purpose, a cluster expansion of the self-energy can be performed.
Ladder-T matrices in $\Sigma(k,iz_\lambda)$%, defined at the Matsubara frequencies $z_\nu= \pi \beta (2 \nu +1)$, 
are calculated by solving the Schrödinger equation derived in \cite{RKKKZ78,ZKKKR78} for $m$ particles in the medium.
 For $m=2$, it reads 
 \begin{eqnarray}
\label{mSGl}
&&(E_1^{\rm qu}+ E_2^{\rm qu}-E^{\rm qu}_{P,\nu} ) \psi_{P,\nu}(1,2) \nonumber \\ &&+[1-f(1)-f(2)] \sum_{1'2'} V^{\rm eff}_{12,1'2'}\psi_{P,\nu}(1',2')=0,
\end{eqnarray}
$f_i(k)=[\exp{(E_k-\mu_i)/k_{\rm B}T}+1]^{-1}$ is the Fermi distribution function. 
 This equation (\ref{mSGl}) contains the screened potential  $V^{\rm eff}_{ij,i'j'}$ (as an approximation for the dynamically screened interaction) and the Pauli blocking term, which describes the exchange term due to antisymmetrisation. 
 In the approximation considered in \cite{RKKKZ78,ZKKKR78},  
the self-energy contains the Debye shift and the Fock shift (exchange term).

The solution of the $m$-particle Schrödinger equation (\ref{mSGl}) generally has a bound state spectrum (ground state and excited states) as well as a continuum solution. 
\GR{ The quantum numbers $P$ denotes the total momentum, and $\nu $ the intrinsic quantum state, in general bound states and scattering states.
 For example, $\nu $ denotes the  ground state when bound states exist, and the excited bound states ($nlms$ for the H atom), but also the continuum ($Elms$ for the H atom, with the energy $E$ of the relative motion.)}

After this treatment of intrinsic energy, we obtain the solution for density as (only two-particle correlations)
\begin{equation}
\label{nTmu}
n(T,\mu_i)= \frac{1}{\Omega_0}\sum_k f_i(E^{\rm qu}_k)+ \frac{1}{\Omega_0}\sum_{P,\nu}g_\nu (E^{\rm qu}_{P,\nu}),
\end{equation}
$g_\nu(E)=[\exp{(E-\mu_2)/k_{\rm B}T}-1]^{-1}$ is the Bose distribution function \GR{ for the two-Fermion system}; $\mu_2$ is the sum of the chemical potentials of the components involved. We repeat that the sum over $\nu$ also includes the continuum of scattering states. 

We briefly outline the derivation of the Beth-Uhlenbeck formula for treating the second term of Eq. (\ref{nTmu}).
For the ideal gas in the classical limiting case, the following applies 
\begin{equation}
\frac{n_i \Lambda_i^3}{2}=e^{\beta \mu^{(0)}_i}=z^{(0)}_i, \,\, \Lambda_i^2 =\frac{2 \pi \hbar^2}{m_i k_{\rm B}T}
\end{equation}
The second term of Eq. (\ref{nTmu}) contains the chemical potentials of both particles in $g_\nu$, so that it is of second order in $z_i^{(0)}$, 
\begin{equation}
\label{eq:2part}
\sum_\nu g_\nu = e^{\beta \mu_i+\beta \mu_j} e^{-\beta \hbar^2 P^2/(2m_i+2m_j)}\hat g_{ij} \sigma_{ij} (T),
\end{equation}
where $ \sigma_{ij} (T)$ is the intrinsic partition function, and $\hat g_{ij}$ a degeneracy factor (spin).
We decompose the sum $\sum_{P,\nu}$ into different channels $c$, which are characterized by the species $i=\{e,p\}$, the spin and the angular momentum, which are conserved quantities of the two-particle system.
Then the sum over these intrinsic quantum numbers yields the degeneracy factor $\hat g_c$, the  center-of-mass momentum is integrated. For the intrinsic partition function $\sigma _c(T)$, Beth and Uhlenbeck derived a quantum statistical expression \cite{BU},
\begin{equation}
\label{sigmaBU0}
\sigma_c(T)=\sum_s e^{-\beta E_{c,s}}+ \int \frac{dE}{\pi} e^{-\beta E} \frac{d}{dE}  \delta_c(E)
\end{equation}
 which contains the sum over the bound states in channel $c$ and the integral over the scattering phase shifts in this channel.
 Using simple algebra, the relations $n_i(T,\mu_j)$, equation (\ref{nTmu}), can be resolved as $\mu_i(T,n_j)$. After the virial expansion with respect to $n_i$, we obtain the second virial coefficient, which contains the intrinsic partition function $\sigma_c(T)$. 
 After integration (\ref{fp(n)}), we obtain the free energy density as a virial expansion, and differentiation yields the virial expansion for the pressure.
 
 With regard to the second virial coefficient, we emphasize the following:
 \begin{enumerate}
 \item
 According to Eq. (\ref{sigmaBU0}), $\sigma_c(T)$ contains a contribution from the bound states and a contribution from the scattered states. 
 In principle, both contributions should be taken into account.
 \item
 The decomposition of the second virial coefficient is not without arbitrariness; integration by parts yields, for example,
 \begin{equation}
\label{sigmaBU}
\sigma_c(T)=\sum_s[ e^{-\beta E_{c,s}}-1]+ \int \frac{dE}{\pi T} e^{-\beta E}  \delta_c(E)
\end{equation}
where Levinson's theorem was used.
\item
 Part of the contribution of the second virial coefficient is shifted to the single-particle contribution when quasi-particles are introduced.
 \end{enumerate}
 In particular, the Hartree-Fock quasiparticle shift contains the lowest order of interaction, so this lowest order term cannot appear in the second virial coefficient \cite{Schmidt90}. 
 Furthermore, the expansion of the Fermi distribution function for the quasiparticles contributes to the second virial coefficient.
 For these reasons, it is not possible to define the amount of bound or free electrons in a plasma without arbitrariness. 
 Only the total second virial coefficient is uniquely defined.

A particular problem arises with  the long-range Coulomb interaction. Standard phase shifts cannot be defined, and divergences occur in the virial expansion with respect to the powers of density. This mathematical problem is solved when screening is taken into account. 
Instead of the expansion in powers of $n$, terms with $n^{1/2}$ occur, so that the expansion in powers of $n$ does not converge.
We give the modified virial expansion in the following section \ref{sec:virial}.

%%%%%%%%%%%%%%%%
\subsection{The Free energy for Coulomb systems}
\label{sec:virial}

For simplicity, we consider the UEG, consisting of electrons ($e$) and a charge compensating homogeneous background, as well as the H plasma, consisting of electrons and protons ($p$). Charge neutrality results in $n_e=n_p=n$.
For the thermodynamic variables temperature $T$ and electron density $n$, the free energy $F(T,\Omega_0,N_i)=\Omega_0 f(T,n)$ is the thermodynamic potential, and all other thermodynamic quantities such as pressure $p(T,n)$ or the mean potential energy $\langle V \rangle =V(T,\Omega_0,N_i) $ are derived from $f(T,n)$:
\begin{eqnarray}
\label{eospv}
p(T,n)&=& n \frac{\partial f(T,n)}{\partial n}-f(T,n) \nonumber \\
\frac{1}{\Omega_0} V(T,n) &=& e^2 \frac{\partial f(T,n)}{\partial (e^2)},
\end{eqnarray}
\GR{ where $e$ is the elementary charge that characterizes the strength of the interaction in Coulomb systems.}

Expressions for the free energy density $f(T,n)$ are derived from quantum statistics using Green's functions, Feynman diagram techniques, and partial summations to describe screening and the formation of bound states \cite{Buch,KKER86}.
In equation (\ref{quEnGl}), the self-energy must include the Debye screening (Montroll-Ward term) to avoid divergent expressions
\begin{equation}
\label{sigmaD}
\langle \Sigma_i^{\rm Debye} \rangle= -\frac{e_i^2 \kappa}{8 \pi \epsilon_0}, \qquad \kappa^2 = \sum_j \frac{e_j^2 n_j}{\epsilon_0 k_{\rm B}T},
\end{equation}
\GR{where $1/\kappa$ is the Debye screening length.}

The virial expansion
for the free energy density  (the components $i$ denote species and spin) is
\begin{eqnarray}
\label{Fvir}
&&\beta f(T,n)=\sum_i [n_i \ln(n_i\Lambda_i^3)-n_i]-F_{\rm Debye}(T)n^{3/2}\nonumber \\ &&
-F_1(T)n^2 \ln n-F_2(T) n^2
-F_3(T)n^{5/2} \ln n-F_4(T) n^{5/2}\nonumber \\ &&+{\cal O}(n^3\ln n),
\end{eqnarray}
see \cite{KKER86}, where also expressions for the lowest virial coefficients $F_i(T)$ up to $F_3(T)$ are given.
Due to the long-range nature of the Coulomb interaction, half-exponents of the density and logarithmic terms occur.  
The analytic form of Eq. (\ref{Fvir})  follows from the evaluation of the Feynman diagrams and partial summations.
The general proof of the analytic behavior cannot be given here.

It is advantageous to introduce dimensionless quantities not only for logarithmic expressions, but also for general calculations.
We use atomic units $T_{\rm Ha}$ for temperature $T$ and $n_{\rm Bohr} = n\,a^3_{\rm B}$ for particle densities, 
so that
\begin{equation}
\label{units}
    T_{\rm Ha}=\frac{1}{315777.1}\, \frac{T}{{\rm K}}, \quad n_{\rm Bohr}= 1.4818471\times 10^{-25} n_{\rm e}\,{\rm cm}^3\,.
\end{equation}

\subsection{Benchmarks for the H plasma EoS}
\label{sec:TD-H}

The thermodynamic properties of hydrogen plasma have been extensively studied. For numerical approaches, see the  review \cite{Bonitz24} and the references cited therein.
The virial expansion of the pressure $p(n,T)$ of hydrogen plasma can be derived from the free energy density (\ref{Fvir}) according to Eq. (\ref{eospv}) and has the form
\begin{eqnarray}
\label{virialexp}
\frac{\beta p}{2 n}&=&A_{\rm ideal}(T)-A_{\rm Debye}(T)n_{\rm Bohr}^{1/2}-A_1(T)n_{\rm Bohr} \ln(n_{\rm Bohr}) \nonumber \\
&&-A_2(T)n_{\rm Bohr}-A_3(T)n_{\rm Bohr}^{3/2} \ln(n_{\rm Bohr})-A_4(T) n_{\rm Bohr}^{3/2}\nonumber \\
&&
+{\cal O}(n_{\rm Bohr} \ln(n_{\rm Bohr})).
\end{eqnarray}
The following  exact expression are known \cite{Ebeling67,Ebeling68,KKER86}
\begin{eqnarray}
\label{virialp}
A_{\rm ideal}(T)&=&1,\nonumber\\
A_{\rm Debye}(T)&=& \frac{(2 \pi)^{1/2}}{3 T_{\rm Ha}^{3/2}},\nonumber\\
 A_1(T)&=& 0,\nonumber\\
 A_2(T)&=&\frac{ \pi}{T_{\rm Ha}^{3/2}} \left\{\left[D(\xi_{ee})-\frac{1}{2} E(\xi_{ee})\right] \right. \nonumber \\ && \left.
 +2 \left(\frac{1}{2}+\frac{m_e}{2m_p}\right)^{3/2}D(\xi_{ep}) \right. \nonumber \\ && \left.+\left(\frac{m_e}{m_p}\right)^{3/2}\left[D(\xi_{pp})-\frac{1}{2} E(\xi_{pp})\right]
\right\}\nonumber \\
&&+\frac{\pi}{6T_{\rm Ha}^3} \ln \left[\frac{4 m_e m_p}{(m_e+m_p)^2}\right]
,\label{A2(T)}\nonumber\\
A_3(T)&=&\frac{(2 \pi^3)^{1/2}}{T_{\rm Ha}^{9/2}},
\end{eqnarray}
with
\begin{eqnarray}
\label{xi}
&&\xi_{ee}=-\frac{1}{T_{\rm Ha}^{1/2}},\,\, \xi_{ep}
%=\left[\frac{2m_em_p}{T_{\rm Ha}(m_e+m_p)}\right]^{1/2}
=\left[\frac{2}{1+\gamma}\right]^{1/2}\frac{1}{T_{\rm Ha}^{1/2}}, \\ && \xi_{pp}
%=-T_{\rm Ha}^{-1/2}m_p^{1/2}
=-\frac{1}{\gamma^{1/2}}\frac{1}{T_{\rm Ha}^{1/2}} ; \,\,
\gamma=\frac{m_e}{m_p}=5.446170226 \times 10^{-4}.\nonumber
\end{eqnarray}

We have the direct term
\begin{eqnarray}
\label{Q}
D(x)&=&-\frac{\pi^{1/2}}{8}x^2-\frac{1}{6}\left(\frac{1}{2} C+\ln(3)-\frac{1}{2}\right)x^3
\nonumber \\&&
+\sqrt{\pi} \sum_{m=4}^\infty\frac{\zeta(m-2)x^m}{2^m \Gamma(m/2+1) },\\
%\end{eqnarray}
%\begin{eqnarray}
 E(x)&=&\frac{\pi^{1/2}}{4}+\frac{1}{2}x+\frac{\pi^{1/2}}{4} \ln(2) x^2\nonumber \\&&
+\sqrt{\pi} \sum_{m=3}^\infty\frac{\zeta(m-1)x^m}{2^m \Gamma(m/2+1)}\left(1-\frac{4}{2^m}\right)
\end{eqnarray}
is the exchange term, $\zeta(m)$ denotes the Riemann zeta function, and $C=0.5772156649\dots$ is Euler's constant.
Other definitions of the direct term, denoted as  $Q(x)$, include an additional linear term $-x/6$ \cite{KKER86}. 
This linear term does not contribute in the case of two-component plasma, but is essential for the UEG, Section \ref{sec:UEG}.
\GR{ For $A_4(T)$, only approximate values are given in \cite{KKER86}, see Eq. (2.52) there. An estimate for $A_4(T)$ can be found in Refs. \cite{AlMaMono,Kahlbaum}. }

In order to obtain convergent results for $D(x)$, Eq. (\ref{Q}),  a large number of terms in the sum must be taken into account. 
This applies in particular to $x=\xi_{ep}>0$,  for which bound states occur and a perturbation expansion does not work.  
An asymptotic expansion in a semi-convergent series is useful for calculating these contributions \cite{KKER86}, 
\begin{eqnarray}
\label{Qsigma}
D(x)&=&2 \pi^{1/2} \Theta(x) \left[\tilde \sigma^{\rm PBL}(x)-\frac{x^2}{8}\right] \nonumber \\ &&
-\frac{1}{6}x^3\left[\ln|x|+2 C+\ln(3)-\frac{11}{6}\right]+\frac{1}{12}x-\frac{1}{60\,x} \nonumber \\ &&+{\cal O}(x^{-3}),
\end{eqnarray}
where $\Theta(x)=1$ if $x>0$, and 0 else. The Planck-Brillouin-Larkin partition function is then given by
\begin{equation}
\label{PBLtilde}
\tilde \sigma^{\rm PBL}(x) =\sum_{s=1}^\infty s^2\left(e^{x^2/(4 s^2)}-1- \frac{x^2}{4 s^2}\right) . 
\end{equation}

This result is interesting. In particular, the contribution of the $e-p$ interaction to $D$, where $\xi_{ep}$ is positive, contains the bound states. Replacing $x^2$ with $\xi_{ep}^2\approx 2/T_{\rm Ha}$ (\ref{xi}), we obtain the following form for the Planck-Brillouin-Larkin partition function 
\begin{equation}
\label{PBL}
\sigma^{\rm PBL}(T) =\sum_{s=1}^\infty s^2\left(e^{1/(2 s^2T_{\rm Ha})}-1- \frac{1}{2 s^2T_{\rm Ha}}\right) 
\end{equation}
(spin factors are treated separately, see Eq. (\ref{eq:2part})).
We can consider this as the convergent part of the bound states to the second virial coefficient.
The term $-1$ already appears in the Beth-Uhlenbeck formula (\ref{sigmaBU}) and avoids jumps in the contribution of the bound state when a bound state merges with the continuum at the Mott point. 
Since the contribution of the scattered state is often neglected, it is to be expected that after this renormalization of the contribution of the bound states, the neglect of the continuum of scattered states is less drastic.
The appearance of the second correction term $-1/(2 s^2T_{\rm Ha})$ is due to the long-range Coulomb interaction.
We must use the generalized Beth-Uhlenbeck expansion (\ref{nTmu}) with the quasiparticle shift (\ref{sigmaD}).
This part of the interaction, which is already taken into account in the quasi-particle contribution, must be subtracted from the second virial coefficient so that this term  $-1/(2 s^2T_{\rm Ha})$ arises and Coulomb divergences are avoided.
Note that we have exact expressions for the second virial coefficient of Coulomb plasmas, so that the different splitting into contributions from bound and scattered states is under control.

The virial expansion can be used as a benchmark for other approaches to calculating the thermodynamic properties of H plasmas.
Accurate results are expected from PIMC simulations \cite{Millitzer,Dornheim2018,PIMC1,B20,Dorn25}, but these suffer from finite size effects and the sign problems.
Systematic results covering the temperature-density plane have recently been published \cite{Filinov23}.
On the one hand, comparison with the above benchmarks allows an assessment of the accuracy of the PIMC data.
On the other hand, it allows an estimation of the applicability of the virial expansion at increasing densities.

\begin{figure}[htp]
\centerline{\includegraphics[width=0.5 \textwidth]{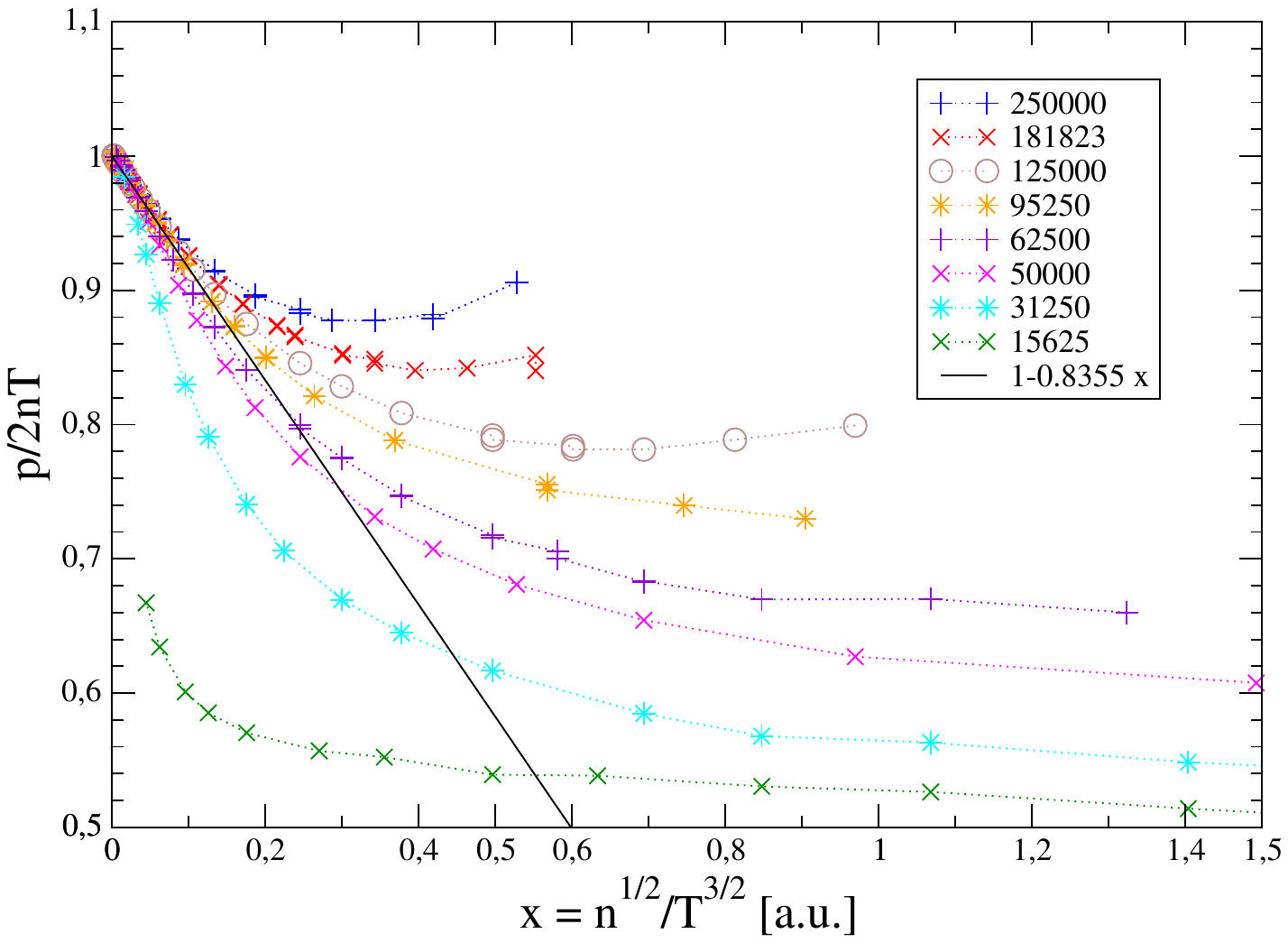}}
\caption{$\beta p/(2 n)$ as function of $n_{\rm Bohr}^{1/2}/T_{\rm Ha}^{3/2}$. PIMC simulations \cite{Filinov23} are shown as a function of $x=n_{\rm Bohr}^{1/2}/T_{\rm Ha}^{3/2}$ for different temperatures in K. The Debye limit $1-x (2 \pi)^{1/2}/3$ is also shown. 
Reproduced with permission from Contrib. Plasma Phys. e70078 (published online 2026).
\label{fig:1a}}
\end{figure}

A detailed analysis was recently carried out \cite{RCER25}; here we summarize only some of the results.
A virial plot of all PIMC simulation data for $\beta p/2n$ as a function of $x=n_{\rm Bohr}^{1/2}/T_{\rm Ha}^{3/2}$ is shown in Fig. \ref{fig:1a}.

The Debye limit $1-x (2 \pi)^{1/2}/3=1-0.8355\, x$ is well reproduced, especially for $ T_{\rm Ha}\ge 1$.
A detailed investigation reveals deviations that limit the accuracy of the published PIMC data to about $10^{-2}$. 
It is interesting to note that the simulations were performed in the critical range $ T_{\rm Ha}\approx 1$, where bound states become important for the thermodynamic properties. 
We see the deviation from the lowest order of the virial expansion, the Debye limit  $1-x (2 \pi)^{1/2}/3$, already at low densities. At the lowest temperature $T = 15625$ K, the Debye limit is not reached for the given data.
As shown in Ref. \cite{RCER25}, the second virial coefficient becomes dominant, and the Saha approximation is a suitable approximation over a wide density range. 

An alternative approach to the standard virial expansion is to start from a generalized Beth-Uhlenbeck approach which takes excited bound states into account. 
For the density dependence of the  quasi-particle shifts (single particle and bound states) as well as the continuum contributions,  a virial expansion  of the few-body self-energies is proposed \cite{RCER25}.

An interesting question is whether we can extract higher virial coefficients from PIMC simulations.
In the virial plot method, we introduce the effective virial coefficient $A_2^{\rm eff}(T,n)$.
As a test, we focus on the second virial coefficient,
since analytical expressions for $A_2(T)$ are known, see Eq. (\ref{virialp}). We use the virial expansion for pressure, Eq. (\ref{virialexp}), and neglect the virial terms $A_m$ with $m\ge 3$. This gives us a virial diagram for an effective second virial term 
\begin{eqnarray} \label{A2eff}
&&A_2^{\rm eff}(T,n_{\rm Bohr})= \left[-\frac{\beta p}{2n} +A_{\rm ideal}(T)-A_{\rm Debye}(T)n_{\rm Bohr}^{1/2}\right. \nonumber \\ && \left. -A_1(T)n_{\rm Bohr} \ln(n_{\rm Bohr})\right] \frac{1}{n_{\rm Bohr}}\nonumber  \\ 
&&=A_2(T)+A_3(T) n_{\rm Bohr}^{1/2} \ln( n_{\rm Bohr})+{\cal O}(n^{1/2}),
\end{eqnarray}
see Eq. (\ref{virialexp}). If we assume that the next virial term $A_3(T)$ provides the leading correction term in the low-density limit, 
a virial plot of $A_2^{\rm eff} (T,n_{\rm Bohr})$ against the abscissa $x_2=n_{\rm Bohr}^{1/2} \ln( n_{\rm Bohr})$ yields  $A_3(T)$ as the slope  at $x_2=0$.
\GR{This was done in Ref. \cite{RCER25}, but the accuracy of the PIMC data is insufficient to obtain a meaningful virial diagram. 
Even the virial diagram for the Debye term shown in Ref. \cite{RCER25} contains errors that are greater than the claimed value of less than 1\%. 
In contrast, highly accurate data are available for the UEG, see Section \ref{sec:UEG}.}

The treatment of the logarithmic terms in the virial expansion requires a special treatment since the argument of the logarithm must be dimensionless, and the division by the logarithm is not possible if the argument takes the value 1. 
In Ref. \cite{RCER25} a method is proposed which circumvents these problems.
We rewrite
\begin{eqnarray}
\label{v2eff1}
A_2^{\rm eff}(T,n)&=&A_2(T)+A_3(T)n_{\rm B}^{1/2} \ln(n_{\rm B})+
A_4(T) n_{\rm B}^{1/2}\nonumber \\&&+{\cal O}(n_{\rm B} \ln(n_{\rm B})\nonumber \\
&=&A_2(T)+A_3(T)n_{\rm B}^{1/2} \ln\left[B_4(T)n_{\rm B}\right]\nonumber \\&&+{\cal O}(n_{\rm B} \ln(n_{\rm B})
\end{eqnarray}
with 
\begin{equation}
\label{A34}
    A_4(T)=A_3(T) \ln [B_4(T)].
\end{equation}
\begin{figure}[h]
\centerline{\includegraphics[width=0.6 \textwidth]{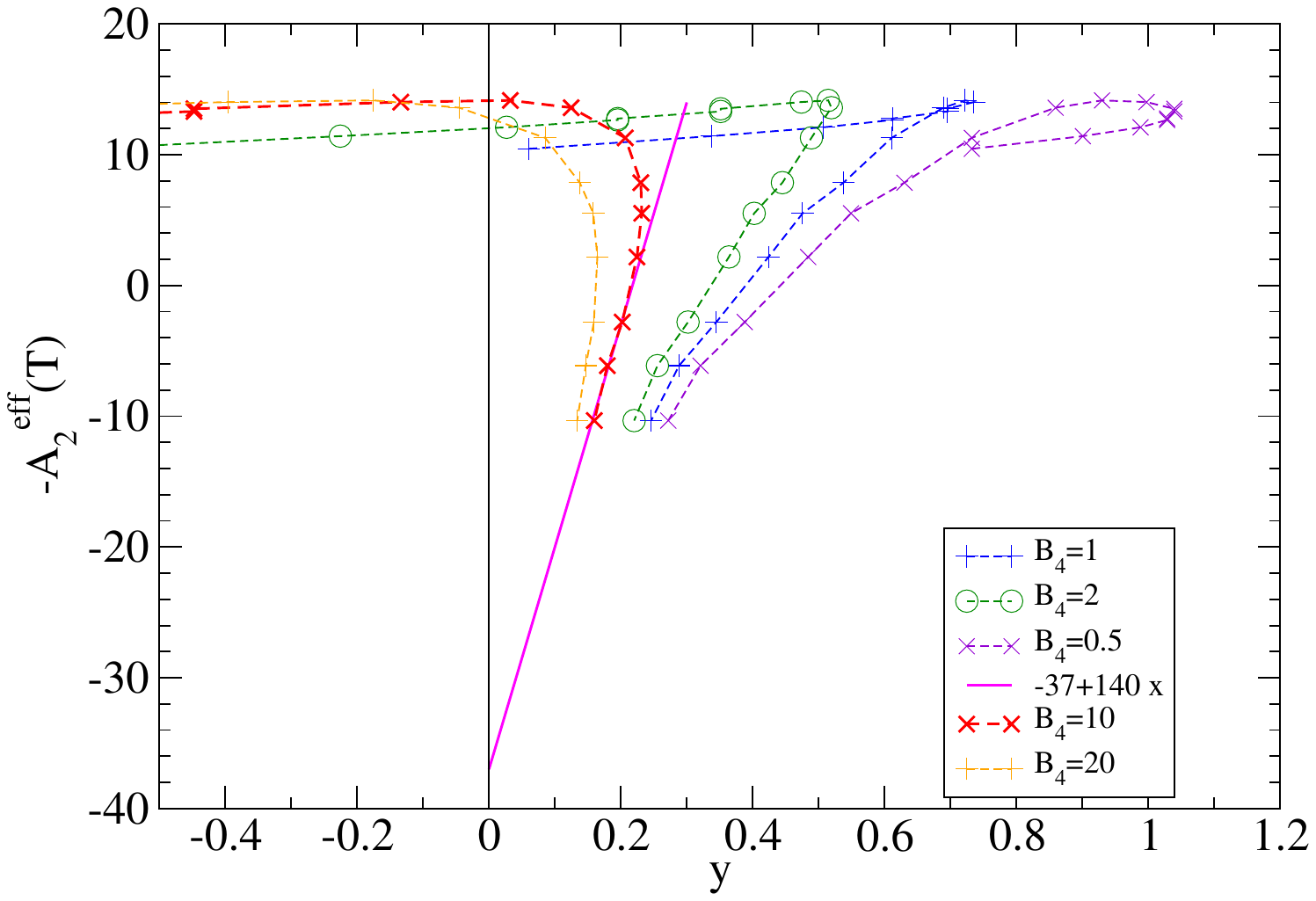}}
\caption{The effective second virial coefficient $A^{\rm eff}_2$ as a function of $y=n_{\rm B}^{1/2} \ln\left[B_4(T)n_{\rm B}\right]$ according to equation (\ref{v2eff1}). The parameter values are $T = 125000$ K, with different values assumed for $B_4$. 
For comparison, the linear extrapolation [neglecting the terms ${\cal O}(n_{\rm B} \ln(n_{\rm B} )$] with $A_2=37.004$ and $A_3 \approx 140$ are also shown.
Reproduced with permission from Contrib. Plasma Phys. e70078 (published online 2026).
\label{fig:3vir125}}
\end{figure}
Thus, in the virial plot, where $A_2^{\rm eff}(T,n)$ is represented as a function of $y=n_{\rm B}^{1/2} \ln\left[B_4(T)n_{\rm B}\right]$, the isotherms should intersect the coordinate $y=0$ at $A_2(T)$ and have a slope of $A_3(T)$. 
The coefficient $B_4(T)$ is determined for the best linear relationship in this virial plot.
The linear pattern is violated when higher virial coefficients become relevant. Note that both $A_2(T)$ and $A_3(T)$ are known for the hydrogen plasma according to Eq. (\ref{A2(T)}).

\GR{This method of virial plots for PIMC simulations $p^{\rm PIMC}(T,n)$ was applied in \cite{RCER25} to derive the virial coefficients $A_3(T), A_4(T)$ from the values $A_2^{\rm eff,PIMC}(T,n)$ according to equation (\ref{v2eff1}).
We use the PIMC results $p=p^{\rm PIMC}$ \cite{Filinov23} and take  $B_4(T)$ in the variable $y$ as a parameter. 
As an example, we show results for the temperature $T=125000$ K, Fig. \ref{fig:3vir125}.
The values for the lowest densities show a strong spread and are discarded.
The known value $A_2=37.004$, equation (\ref{virialp}), is used for the linear extrapolation.}

\GR{It is obvious that this method for extracting virial coefficients requires high accuracy of the calculated data $p^{\rm PIMC}(T,n)$. 
We analyze the difference of large numbers, since the lower virial coefficients, such as the Debye term, dominate the limit value of the pressure at low density. 
The best choice for $B_4$ is the graph in which the linear relationship is best realized.
Taking into account the various graphs in Fig. \ref{fig:3vir125}, the PIMC data follow a linear relationship as far as possible
for $B_4 \approx 10$.
 The corresponding value $A_3 \approx 140$ (\ref{A34}) is smaller than the theoretical value 509.744 according to equation (\ref{virialp}).
 A value for $B_4$ greater than 10 is possible, but for a better determination of $B_4$, we need accurate PIMC data for low densities.
We only show the procedure here, but the extracted values for the higher virial coefficients are very uncertain due to the strong scattering of the PIMC data at low densities.
Further details can be found in Ref. \cite{RCER25}.}

\subsection{Benchmarks for the thermodynamics of the UEG}
\label{sec:UEG}

The uniform electron gas (UEG) is a simpler model in which only electrons (two spin directions) are considered.
The repulsive interaction allows for a simpler treatment of the second virial coefficient; no bound states arise. Instead of the free energy, see \cite{KKER86}, we consider the virial expansion for the density of the mean potential energy (\ref{eospv}), which is closely related to the free energy density. 
Accurate PIMC simulations have been performed  for the UEG \cite{TD}.

We use atomic (Hartree) units so that $V_{\rm Ha} = V/(27.2114$ eV), and perform the virial expansion for the specific mean potential energy $v=V_{\rm Ha} /N$ as
\begin{eqnarray}
\label{virialv}
&&v(T,n)=v_{\rm Debye}(T) n_{\rm Bohr}^{1/2}+v_1(T) n_{\rm Bohr} \ln\left(\kappa^2 \lambda^2\right)
\nonumber \\ &&+v_2(T) n_{\rm Bohr}+v_3(T)n_{\rm Bohr}^{3/2} \ln\left(\kappa^2 \lambda^2\right)+v_4(T) n_{\rm Bohr}^{3/2}
\nonumber \\&& +{\cal O}(n^2 \ln(n)),
\end{eqnarray}
where $\kappa^2 \lambda^2=4 \pi n_{\rm Bohr}/T_{\rm Ha}^2$. Using the virial expansion for the free energy \cite{KKER86} we find with (\ref{eospv})
\begin{eqnarray}
v_{\rm Debye}(T)&=&-\frac{\sqrt{\pi}}{T_{\rm Ha}^{1/2}}, \nonumber  \\ v_1(T)&=&-\frac{\pi}{2 T_{\rm Ha}^2},\nonumber\\
v_2(T)&=&-\frac{\pi}{T_{\rm Ha}}\left[\frac{1}{2}-\frac{\sqrt{\pi}}{2}(1+\ln(2))\frac{1}{T_{\rm Ha}^{1/2}}\right. \nonumber \\ && \left.+\left(\frac{C}{2}+\ln(3)-\frac{1}{3}+\frac{\pi^2}{24} \right) \frac{1}{T_{\rm Ha}}\right.\nonumber \\
&&\left. - \sqrt{\pi}\sum_{m=4}^\infty\frac{m}{2^m \Gamma(m/2+1)}\left(\frac{-1}{T_{\rm Ha}^{1/2}}\right)^{m-1}  \right. \nonumber \\ && \left. \times[2 \zeta(m-2)-(1-4/2^m)\zeta(m-1)]\right],\nonumber \\
v_3(T)&=-&\frac{3 \pi^{3/2}}{2 T_{\rm Ha}^{7/2}}.
\label{v0123}
\end{eqnarray}
Note that for the UEG, the virial coefficient $v_1(T)$ does not disappear. 
In contrast to \cite{KKER86}, the linear term $-x/6$ does not appear in Eq. (\ref{Q}) for $D(x)$, the direct term of the second virial coefficient.

This exact result for the virial expansion of the UEG equation of state can be compared with the numerical results from PIMC simulations \cite{TD}, which claim to provide highly accurate results for the EoS over a wide density range.
We collect the numerical results given in \cite{TD} and \cite{R24} in Appendix \ref{app:UEG}.
In contrast to the PIMC data for the H plasma, we have no isotherms.
Interpolation between these data is obtained from the formula $v^{\rm GDSMFB}$ given in Ref. \cite{GDSMFB17}.
This formula has an inaccuracy of about 0.1 \%.
Within these errors, PIMC calculations $v^{\rm PIMC}$ are replaced by the interpolation formula $v^{\rm GDSMFB}$.

\begin{figure}[t]
\centerline{\includegraphics[width=0.4 \textwidth]{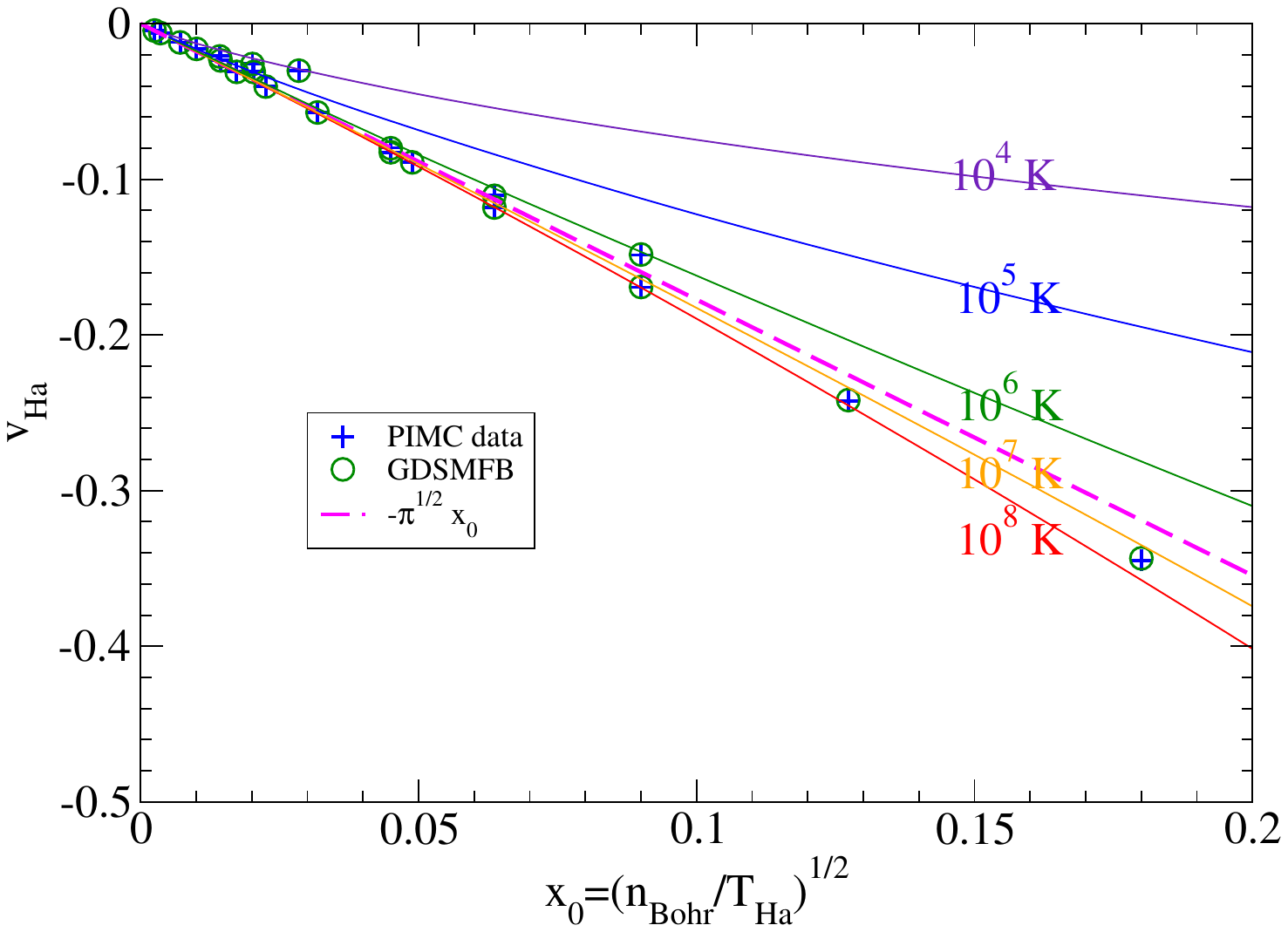}}
\caption{Potential energy $v(T,n)$ as a function of $x_0=(n_{\rm Bohr}/T_{\rm Ha})^{1/2}$. 
PIMC simulation data according Tab. \ref{Tab:v}. 
Also shown are the interpolation result $v^{\rm GDSMFB}(T,n)$ according  \cite{GDSMFB17} and the Debye term 
$v_{\rm Debye}(T) n_{\rm Bohr}^{1/2}=-\pi^{1/2} x_0$ (dashed line).
Isotherms are shown for $v(T,n)$, Eq. (\ref{virialv}), neglecting $v_4$ and higher order virial coefficients.
  (Atomic units are used.)\label{fig:1}}
\end{figure}

To demonstrate the Debye term, we show the expression $v(T,n)$ as a function of $x_0=(n_{\rm Bohr}/T_{\rm Ha})^{1/2}$ in Fig. \ref{fig:1}.
The Debye term $v_{\rm Debye}(T) n_{\rm Bohr}^{1/2}=-\pi^{1/2} x_0$ is also shown.
For $v(T,n)$ we can take the simulation values $v^{\rm PIMC}(T,n)$. 
Alternatively, we can also take the approximation $v^{\rm GDSMFB}(T,n)$.
We see:
\begin{enumerate}
\item
The values $v^{\rm PIMC}(T,n)$ coincide with $v^{\rm GDSMFB}(T,n)$, the errors are smaller than the symbols.
\item
The Debye limit $-\pi^{1/2} x_0$ is well reproduced for small values of $x_0$. 
Deviations for $T \le 10^6$ K are seen for $x_0 <0.05$. These deviations indicate that higher orders of the virial expansion must be considered.
\item
The accuracy of the PIMC data is very high, the scatter around the Debye limit near $x_0 \to 0$ is small.
\end{enumerate}

We extract the virial coefficients from the simulation data using the virial plots. 
We demonstrate this method considering examples where the results are known.
Since we have no isotherms for the PIMC simulations, we take two temperatures where PIMC simulations are performed for three densities, see Tab. \ref{Tab:4}, \# 16, 11, 18 for $T_{\rm Ha}=0.589307$, and \# 17, 12, 19 for $T_{\rm Ha}=0.294653$.

\begin{table}[h]
\begin{center}
\hspace{0.5cm}
 \begin{tabular}{|c|c|c|c|c|}
\hline
$T_{\rm Ha}$ &$v_{\rm Debye}(T)$& $v_1(T)$ & $v_2(T)$ & $v_3(T)$ \\ 
\hline
0.589307 & -2.30889 & -4.52311 & -10.3561 & -53.1643 \\
0.294653 & -3.86527 & -18.0924 & -54.181 & -601.489 \\
\hline
 \end{tabular}
\caption{Virial coefficients for the UEG, see Eq. (\ref{v0123}) .}
\label{Tab:v}
\end{center}
\end{table}

For the systematics of demonstration, we start with the simplest one which is trivial. 
For the lowest virial coefficient, we plot 
\begin{eqnarray}
\label{vDebeff}
v^{\rm eff}_{\rm Debye}(T,n)&=&v(T,n)/n_{\rm Bohr}^{1/2} \\
&=& v_{\rm Debye}(T)+v_1(T)  y_0 +v_2(T) n_{\rm Bohr}^{1/2}\nonumber \\
&&+{\cal O}(n \ln(n))\nonumber 
\end{eqnarray}
as a function of $y_0=n_{\rm Bohr}^{1/2} \ln\left(4 \pi n_{\rm Bohr}/T_{\rm Ha}^2\right)$.
We have the result $ v_{\rm Debye}(T) = \lim_{n \to 0} v_{\rm Debye}^{\rm eff}(T,n)$ and the slope there gives $v_1(T)$, see Tab. \ref{Tab:v}.

\begin{figure}[t]
\centerline{\includegraphics[width=0.4 \textwidth]{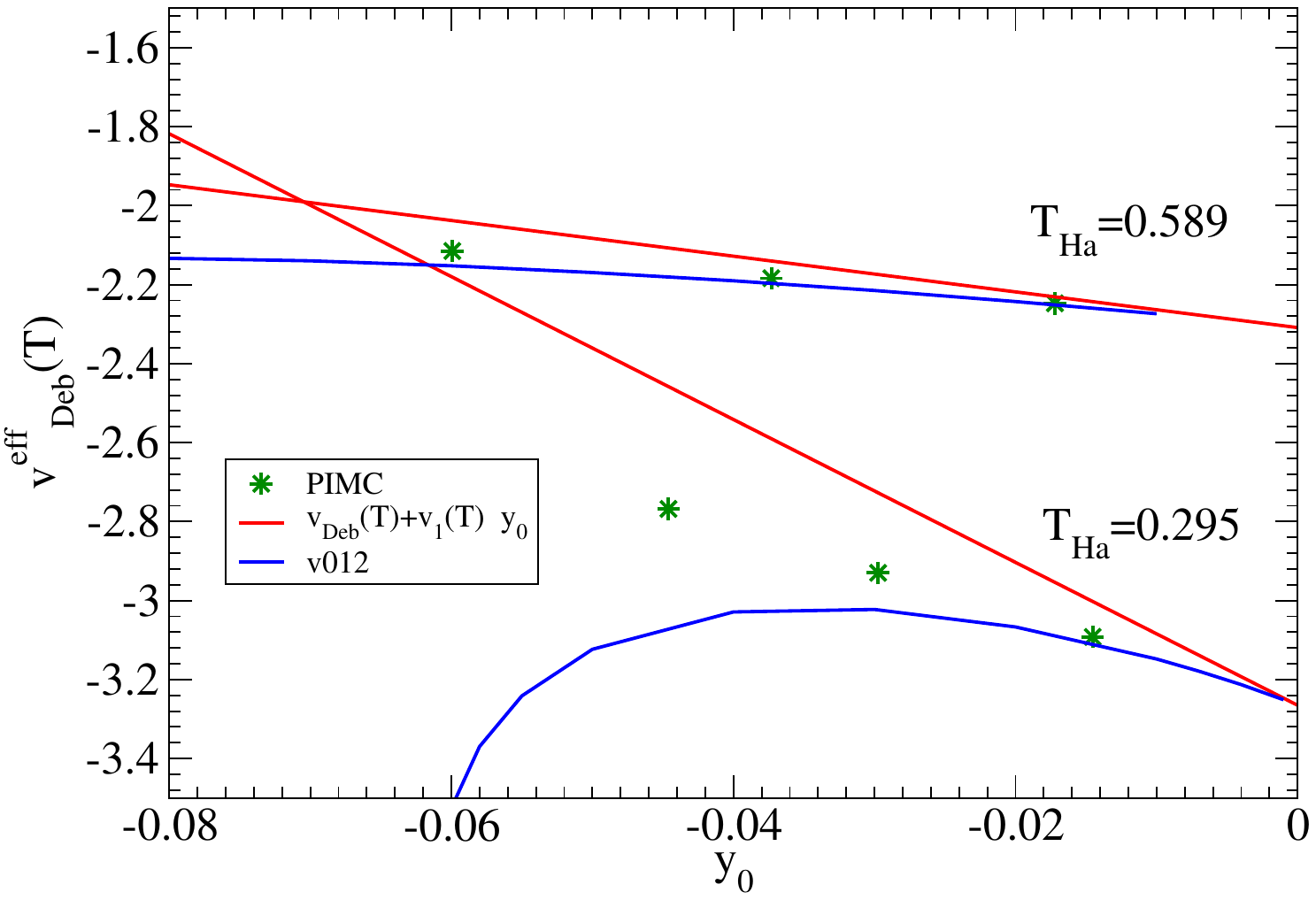}}
\caption{The effective Debye virial coefficient $v^{\rm eff}_{\rm Debye}(T,n)$ (Ha units) as a function of $y_0=n_{\rm Bohr}^{1/2}\ln(4 \pi n_{\rm Bohr}/T^2_{\rm Ha})$. Also shown are  $v_{\rm Debye} (T)+v_1(T) y_0$ and  $v012=v_{\rm Debye} (T)+v_1(T) y_0+v_2(T) n_{\rm Bohr}^{1/2}$. Isotherms for $T_{\rm Ha}=0.589307$ and $T_{\rm Ha}=0.589307/2$, $v^{\rm PIMC}(T,n)$ according Tab. \ref{Tab:4}.
  \label{fig:vDeb}}
\end{figure}

Values are shown in Fig. \ref{fig:vDeb}.
% We see that the interpolation formula do not match the correct behavior.\\
If we replace in Eq. (\ref{vDebeff}) $v(T,n)$ by the calculated values $v^{\rm PIMC}(T,n)$, see Tab. \ref{Tab:4}, we see that $|y_0|$ is not sufficiently small for these data to perform the limit $|y_0| \to 0$. 
In the range where the PIMC simulations are performed, the next virial coefficient $v_2(T)$ is already important so that the linear range  is not reached where  $v_{\rm Debye} (T)+v_1(T) y_0$ is a good approximation.
Nevertheless, the effective values $v^{\rm eff}_{\rm Debye}(T,n)$ are already close to $v_{\rm Debye} (T)$.

To extract the next virial coefficient $v_1(T)$ from data, we consider the quantity
\begin{eqnarray}
\label{v1eff}
v_1^{\rm eff}(T,n)&=&\frac{v(T,n)-v_{\rm Debye}(T) n_{\rm Bohr}^{1/2}}{n_{\rm Bohr} \ln(4 \pi n_{\rm Bohr}/T_{\rm Ha}^2)}\\  &=&v_1(T)+v_2(T)  x_1+v_3(T) n_{\rm Bohr}^{1/2} \nonumber \\ &&+{\cal O}(n^{1/2}/\ln(n))\nonumber 
\end{eqnarray}
as a function of $x_1=1/ \ln\left(4 \pi n_{\rm Bohr}/T_{\rm Ha}^2\right)$.
We have the result $v_1(T) = \lim_{n \to 0} v_1^{\rm eff}(T,n)$ and the slope there gives $v_2(T)$, see Tab. \ref{Tab:v}.

\begin{figure}[h]
\centerline{\includegraphics[width=0.4 \textwidth]{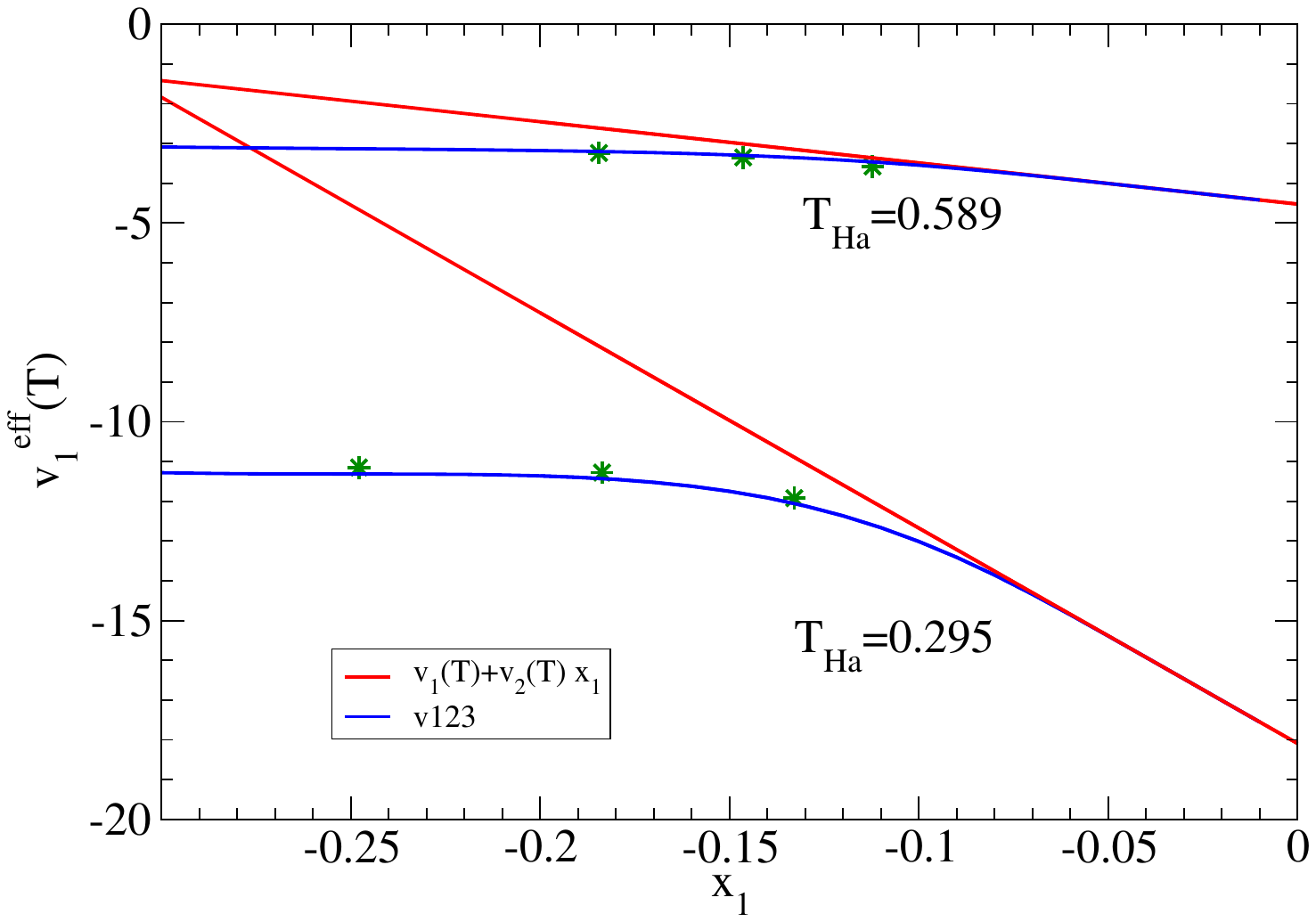}}
\caption{The effective first virial coefficient $v^{\rm eff}_1(T,n)$ as a function of $x_1=1/\ln(4 \pi n_{\rm Bohr}/T^2_{\rm Ha})$. Also shown are  $v_1 (T)+v_2(T) x_1$ and  $v_1 (T)+v_2(T) x_1+v_3(T) n_{\rm Bohr}^{1/2}$. Isotherms for $T_{\rm Ha}=0.589307$ and $T_{\rm Ha}=0.589307/2$, $v^{\rm PIMC}(T,n)$ according Tab. \ref{Tab:4}.
  \label{fig:v1}}
\end{figure}

Values are shown in Fig. \ref{fig:v1}.
% We see that the interpolation formula do not match the correct behavior.\\
If we replace in Eq. (\ref{v1eff}) $v(T,n)$ by the calculated values $v^{\rm PIMC}(T,n)$, see Tab. \ref{Tab:4}, we see that $|x_1|$ is not sufficiently small for these data to perform the limit $|x_1| \to 0$. 
In the range where the PIMC simulations are performed, the next virial coefficient $v_3(T)$ is already important so that the linear range  is not reached where $v_1 (T)+v_2(T) x_1$ is a good approximation.
$v_1(T)$ gives only a small correction \cite{R24} so that it is not easy to separate it.
The series expansion is problematic since $\ln(4 \pi n_{\rm Bohr}/T_{\rm Ha}^2)$ can take the value zero so that the convergence of the series expansion  for $v_1^{\rm eff}(T,n)$ is limited.

To extract the next virial coefficient $v_2(T)$ from data, we consider the quantity
\begin{eqnarray}
\label{v2eff}
v_2^{\rm eff}(T,n)&=&\left[v(T,n)-v_{\rm debye}(T)n_{\rm Bohr}^{1/2}\right. \nonumber \\
&&\left.-v_1(T)n \ln\left(4 \pi n_{\rm Bohr}/T_{\rm Ha}^2\right)\right]/n_{\rm Bohr} \nonumber \\
&=&v_2(T)+v_3(T) x_2+{\cal O}[n^{1/2}]
\end{eqnarray}
as a function of $x_2=n_{\rm Bohr}^{1/2} \ln\left(4 \pi n_{\rm Bohr}/T_{\rm Ha}^2\right)$.
We have the result $v_2(T) = \lim_{n \to 0} v_2^{\rm eff}(T,n)$ and the slope there gives $v_3(T)$, see Tab. \ref{Tab:v}. 
%The density dependence of $v_2^{\rm eff}(T,n)$ in the low-density limit is given according Eq. (\ref{virialexp}) as
%\begin{equation}
%\label{v2eff1}
%v_2^{\rm eff}(T,n)=v_2(T)+v_3(T)n^{1/2} \ln(4 \pi n/T^2)+{\cal O}[n^{1/2}].
%\end{equation}

\begin{figure}[h]
\centerline{\includegraphics[width=0.4 \textwidth]{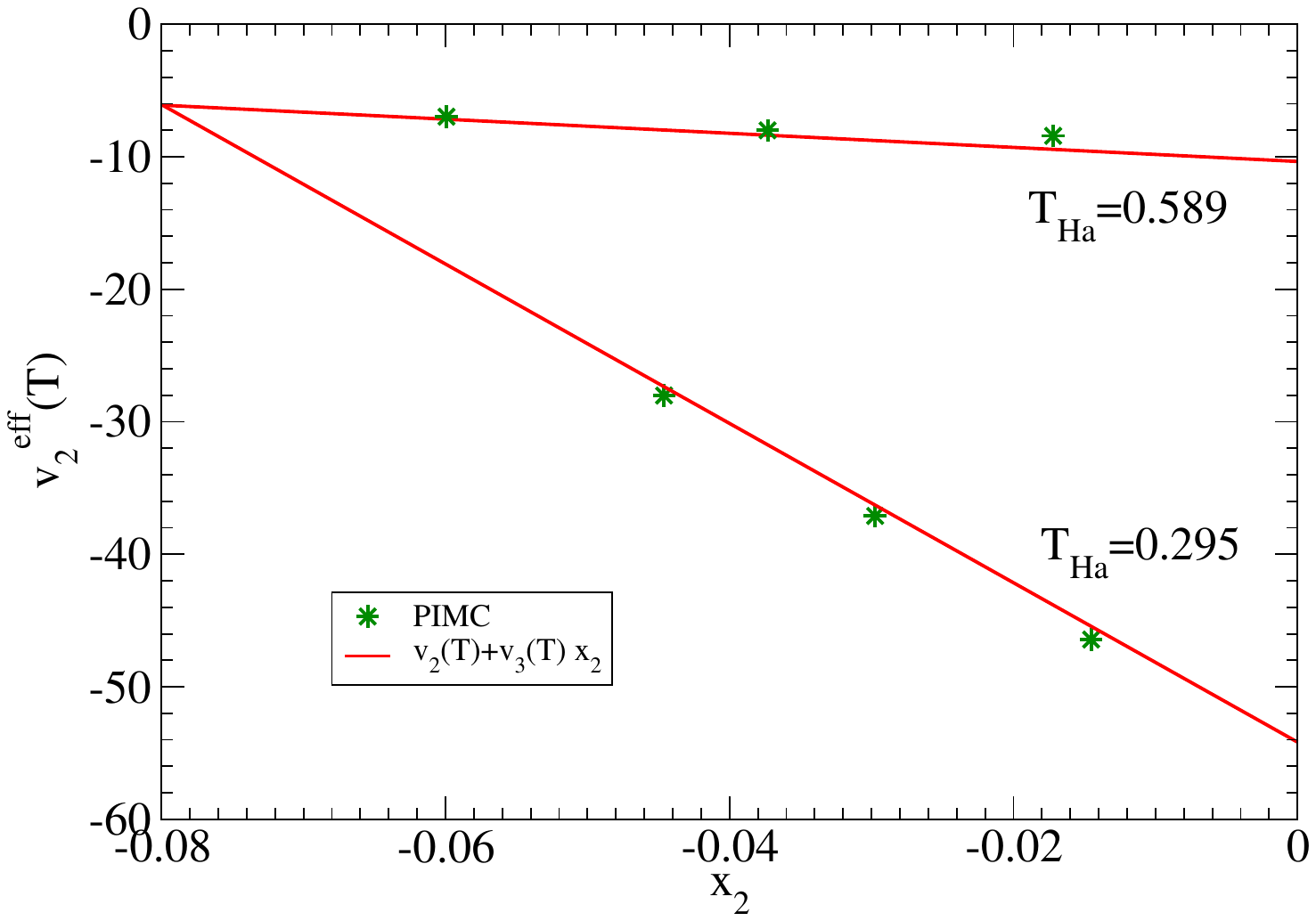}}
\caption{The effective second virial coefficient $v^{\rm eff}_2(T,n)$ as a function of $x_2=n_{\rm Bohr}^{1/2} \ln(4 \pi n_{\rm Bohr}/T^2_{\rm Ha})$. Also shown are  $v_2 (T)+v_3(T) x_2$. Isotherms for $T_{\rm Ha}=0.589307$ and $T_{\rm Ha}=0.589307/2$, $v^{\rm PIMC}(T,n)$ according Tab. \ref{Tab:4}.
  \label{fig:v2}}
\end{figure}

Values are shown in Fig. \ref{fig:v2}.
% We see that the interpolation formula do not match the correct behavior.\\
If we replace in Eq. (\ref{v2eff}) $v(T,n)$ by the calculated values $v^{\rm PIMC}(T,n)$, see Tab. \ref{Tab:4}, we see the surprising result that $|x_2|$ is sufficiently small for these data to perform the limit $|x_1| \to 0$. The small scatter of the data around  $v_2 (T)+v_3(T) x_2$ indicates the high accuracy of the PIMC simulations.

\GR{ An interesting question refers to the convergence of the virial expansion, i.e. how many term must be considered for a given accuracy?
A possible estimate would be the ratio of successive virial coefficients. 
The problem is that the determination of the virial coefficients is not unique since the logarithmic terms contain a factor which contributes to the next virial coefficient, see Eq. (\ref{v2eff1}).
A calculation of the contribution of different virial terms was shown in \cite{R24}. 
Instructive are also the virial plots, Figs. \ref{fig:vDeb} - \ref{fig:v2} where the deviations from a low-order virial expansion is shown.}

A treatment of the second virial coefficient $v_2(T)$ is given in \cite{R24}.
In particular, the possibility to extract the next virial coefficient $v_4(T)$ from PIMC simulation data has been discussed. It has been found that the current accuracy of the PIMC simulations is not sufficient to extract values for the higher virial coefficients.

In Ref. \cite{TD}, a virial plot was presented to investigate the high-temperature behavior of the second virial coefficient $v_2(T)$.
\begin{eqnarray}
&&-\frac{T_{\rm Ha}}{\pi}v_2^{\rm eff}(T,n)
=\frac{1}{2}-\frac{\sqrt{\pi}}{2} (1+\ln(2))
\frac{1}{T_{\rm Ha}^{1/2}}\nonumber \\ &&+\left(\frac{C}{2}+\ln(3)-\frac{1}{3}
+\frac{\pi^2}{24}\right)\frac{1}{T_{\rm Ha}}\nonumber \\ &&
+{\cal O}(T^{-3/2})+{\cal O}(n^{1/2}\ln(n)).
\label{v2vir}
\end{eqnarray}
The plot of this quantity as a function of $\xi=-T_{\rm Ha}^{-1/2}$ should give at $\xi \to 0$ the value 1/2. 
Using the calculated PIMC data, $v(T,n) \to v^{\rm PIMC}(T,n)$ in Eq. (\ref{v2eff}), this result has been confirmed.
Another result for the second virial coefficient, which contains a linear term in the direct contribution ($Q$) to  the equation of state of plasmas, is not confirmed, see Kraeft et al.  \cite{KKR15}.

This point has been debated further in Ref. \cite{Ebel25} assuming an own
definition of the UEG which is based on a limit of a uniform two-component system of charges, electrons with charge $–e$ with
density $n_e$ and particles with opposite charges $+Ze$ with density $n_i$ satisfying the neutrality condition
$e n_e = Zen_i$; it is assumed that the UEG limit is then defined by
$Z \to 0$ so that  $n_i\to \infty$. 
\GR{For any charge-neutral two-component plasma, it can be shown that a linear term $-\xi/6$ in the direct term $D(\xi)$, equation (\ref{virialp}), is exactly canceled out, see also equation (2.52) in Ref. \cite{KKER86}.
Therefore, the conclusion of \cite{Ebel25} is that the use of this term $-\xi/6$ has no influence on the properties of the two-component system.
However, other properties such as pressure are determined by both components of the two-component plasma and diverge for $n_i\to \infty$.
The two-component plasma cannot be used to define the UEG because the Hamiltonian operator has a different form.}

\GR{The UEG is a single-component plasma that moves in an external potential of a structureless background that is homogeneously charged, so that the Hartree term of the electron self-energy is compensated. This external potential has no dynamic degrees of freedom and does not contribute to the kinetic energy. A corresponding definition of the Hamilton operator is also required when performing PIMC simulations.}

\subsection{Open problems}
\begin{enumerate}
\item
The author is not aware of any rigorous proof of the analytical form in which thermodynamic functions such as free energy density depend on the plasma parameters $T,n$.
Within the framework of the Green's function approach, special versions of virial expansions are obtained after performing partial summations, see equation (\ref{Fvir}).
PIMC simulations confirm these analytical dependencies.

\item
PIMC simulations confirm the values of virial expansion. In principle, numerical results for higher-order virial coefficients can be obtained. 
Virial plots can be used to extrapolate values for the virial coefficients.
The fourth virial coefficient, for example, is of interest.

\item
Standard virial expansions have a limited range of application when bound states occur in the plasma.
Therefore, an alternative approach is proposed in which the virial expansion is performed for the self-energies of the single quasi-particle and the bound states. In addition, the contribution of the continuum must be included in a consistent manner to avoid double counting.

\item 
Accurate PIMC simulations are required to extract virial coefficients. To do this, the size problem must be solved and the sign problem addressed.
The comparison with the virial  expansions as a benchmark can be used to estimate the accuracy of the simulations.
%Highly accurate data is available for the UEG, but not currently for the H plasma.
High-precision ab initio path integral Monte Carlo simulations for the direct estimation of the free energy of the uniform electron gas and hydrogen can be obtained from \cite{Dornheim25}, where a comparison with virial expansions would be of interest.

\item
There are PIMC simulations for isotherms for H plasmas, but the accuracy should be improved. It would be interesting to have more data for isotherms for the UEG as well.
It would also be interesting to expand the range of parameter values in order to obtain a better systematics and boundary cases.

\item
Virial expansions are useful for providing results in the low-density range, where PIMC simulations become very expensive.
Together with PIMC data, they can be used to derive interpolation formulas that provide correct limit values, especially at low densities.

\end{enumerate}

\section{Transport coefficients}
\label{sec:transport}

 Linear response theory provides exact expressions for the transport coefficients in the form of equilibrium correlation functions (fluctuation-dissipation theorem).
 The general approach can be found in textbooks, see \cite{Zubarev,Buch}.
 As an example, we quote the Kubo formula \cite{Kubo1957}, which determines the direct current (dc) conductivity $\sigma_{\rm dc}(T,n)$ through the current-current correlation function,
 \begin{equation}
 \label{Kubo}
\sigma_{\rm dc}(T,n)=\frac{e^2 \beta}{3m^2 \Omega_0} \int_{-\infty}^0 dt e^{\delta t}\int_0^1d \lambda \langle {\bf P}\cdot {\bf P}(t+i\hbar \beta \lambda) \rangle,
 \end{equation}
$\lim \delta \to 0$, 
with the total momentum of the electrons  ${\bf P}=\sum_k\hbar {\bf k} a^+_ka^{}_k$.

The perturbation theory and the Green's function method cannot be directly applied to this Kubo formula, since the conductivity in the zeroth order of interaction is infinite and requires special treatment \cite{Ropke18}.
However, for the reciprocal of the DC conductivity, the resistivity, a perturbative Green's function approach is possible within the generalized linear response theory \cite{Roep88,RR89,Redmer97}.
We introduce the dimensionless resistivity
\begin{equation}
\rho^*(T,n)=\frac{(k_BT)^{3/2}(4 \pi \epsilon_0)^2}{e^2 m_e^{1/2}} \frac{1}{\sigma(T,n)}.
\end{equation}
The exact expression for conductivity (\ref{Kubo}) is replaced by the autocorrelation function of the occupation number operators $\hat n_k= a^+_ka^{}_k$ within the framework of generalized linear response theory.
This reflects the generalization of the Boltzmann collision term to dense systems, and a solution to the Boltzmann equation is obtained with moments ${\bf P}_m=\sum_k\hbar {\bf k} (\beta E_k)^m a^+_ka^{}_k$ of the distribution function \cite{Redmer97,Reinholz12}. 
In line with a variation approach, the solution converges quickly with the number $L$ of moments used.

Linear response theory describes the reaction of the system to an external field in the lowest order of the strength of this perturbing field. 
Within the framework of linear response, we can decompose any external field into components with a specific wave number $\bf q$ and frequency $\omega$. 
The reaction to an electric field yields the dielectric function $\epsilon({\bf q}, \omega)$, which is related to the density-density autocorrelation function $\langle \hat \rho_q \hat{\rho}_{-q}(t+i\hbar \beta \lambda) \rangle$, where $\hat \rho_q=\sum_k  a^+_{k+q}a^{}_k$ is the Fourier component of the position-dependent density \cite{Reinholz05}.

\subsection{Benchmarks for the H plasma conductivity}
\label{sec:sigma}

Exact benchmarks for the resistivity of hydrogen plasma can be provided in the form of virial expansions
\begin{eqnarray}
\rho^*(T,n)&=&\rho_1(T) \ln (\Theta/\Gamma)+\rho_2(T)\nonumber \\ &&+\rho_3(T) n_{\rm Bohr}^{1/2} \ln (\Theta/\Gamma)+{\cal O}(n_{\rm Bohr}^{1/2}).
\end{eqnarray}
For the dimensionless logarithmic term, we used the Born parameter 
\begin{eqnarray}
\label{Gamma}
\frac{\Theta}{\Gamma}&=& \frac{2^{1/3}}{3^{1/3} \pi^{5/3}}\frac{T_{\rm Ha}^2}{n_{\rm Bohr}},\nonumber \\ 
\Gamma&=&\frac{e^2 \beta}{4 \pi \epsilon_0}\left(\frac{4 \pi}{3}n\right)^{1/3}, \,\,\,\, \Theta=\frac{2 m_e}{\beta \hbar^2}(3 \pi^2n)^{-2/3}
\end{eqnarray}
are the well-known plasma parameters, describing the coupling strength and the degeneracy of the electrons. 
As seen in the thermodynamic functions of Coulomb systems (\ref{Fvir}), logarithmic terms and fractional exponents of density appear.

Compared to thermodynamic properties, little is known about the virial coefficients $\rho_i(T)$, as the evaluation of transport coefficients using the method of Green's functions is more complex. 
We have \cite{Spitzer53}, \cite{Karachtanov16}
\begin{eqnarray}
\label{benchsigma}
\rho_1(T)=\rho_1^{\rm Spitzer}&=&0.846024, \nonumber\\
\lim_{T \to \infty} \rho_2(T)& =&0.4917.
\end{eqnarray}
\GR{A lot of work has been done to determine the temperature dependence $\rho_2(T)$, but only approximate values are available, see \cite{R23,pop24} and other references cited therein.
For example, the following interpolation formula was given with $T_{\rm eV}=k_{\rm B}T/{\rm eV}$
\begin{equation}
\rho_2(T) \approx 0.4917.+0.846 \ln \left[\frac{1+8.492/T_{\rm eV}}{1+25.83/T_{\rm eV}+167.2/T_{\rm eV}^2}\right].
\end{equation}
Better results are expected from a systematic quantum statistical approach evaluating two-particle T-matrices. 
An alternative would be to perform high-accurate PIMC simulations for correlation functions such as the Kubo formula (\ref{Kubo}).}

\GR{ A contribution to the third virial coefficient $\rho_3(T)$ is related to the Debye-Onsager relaxation process and describes the deformation of the Debye sphere around a charged particle in the external field \cite{Roep88}.
For classical plasma ($ \Theta \gg 1 $), the expression
\begin{eqnarray}
&&\rho^*(T,n)=0.846024\nonumber \\ 
&& \times \left[\ln \Gamma^{-3}+2.248+0.239 \Gamma^{3/2} \ln  \Gamma^{-3}+\dots \right]
\end{eqnarray}
with $\Gamma= (4 \pi n_{\rm Bohr}/3)^{1/3}/T_{\rm Ha}$, Eq.~(\ref{Gamma}), was given. The last term is the contribution of the Debye-Onsager relaxation process. 
The problems in calculating the higher virial coefficients are the consideration of dynamic screening and strong collisions. 
In addition, the formation of bound states must be taken into account below $T_{\rm Ha} \sim 1$, and  at higher densities, degeneration effects (Pauli blocking) occur.
In this case, too, PIMC simulations are of great importance, similar to the thermodynamic properties discussed in Section \ref{sec:TD-H} and \cite{RCER25}, where the formation of bound states is discussed.}

There are approximate expressions \GR{ for resistivity or its reciprocal, the dc conductivity}, but these lead to an incorrect virial expansion \cite{Roep88}.
The Ziman formula, which is related to the force-force correlation function, can be applied to highly degenerate plasmas.
It corresponds to using only the lowest moment of the distribution function, $L=1$, and yields the incorrect limit value $\rho_1^{\rm Ziman}=(8 \pi)^{1/2}/3=1.67109$.
Another approximation is the Lorentz plasma, in which $e - e$ collisions are neglected. 
The relaxation time approach for elastic collisions can be applied, yielding the incorrect limit $\rho_1^{\rm Lorentz}=(2 \pi^3)^{1/2}/16=0.492126$.
This approximation is often used \cite{LeeMore84}, and the neglect of $e-e$ collisions is less sensitive for ions heavier than hydrogen.
In our calculations, we already achieve good results with the two-moment approximation $L=2$, and a fast convergence is observed when a large number of moments of the electron distribution function are taken into account \cite{Redmer97}.
Various approaches to calculating the dc conductivity of H plasma \cite{Grabowski20,Stanek2024} do not reproduce the benchmark value (\ref{benchsigma}).

In this article, we focus on the comparison with numerical simulations.
For this purpose, we introduce a virial plot to extract the virial coefficients from the data.
We introduce the Born variable 
\begin{equation}
x_{\rm Born}= \frac{1}{\ln(\Theta/\Gamma)}
\end{equation}
and plot 
\begin{equation}
\tilde{\rho}(T,x)=\rho^*x=\rho_1(T)+\rho_2(T) x + \dots
\end{equation}

\begin{figure}[t]
\centerline{\includegraphics[width=0.5 \textwidth]{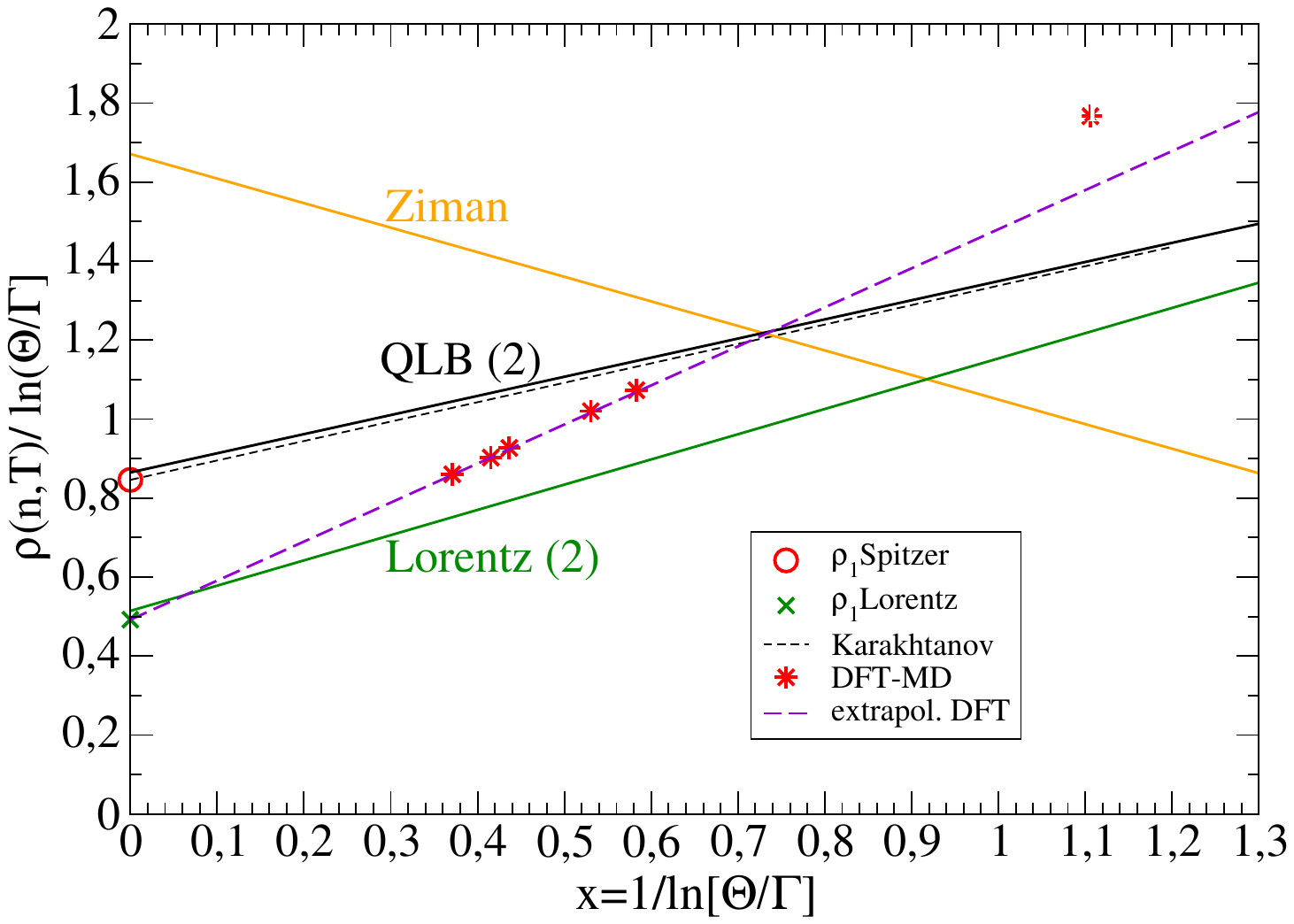}}
\caption{Virial plot for the resistivity.
The benchmark (\ref{benchsigma}) is shown (Karakhtanov \cite{Karachtanov16}) as a function of $x=x_{\rm Born}=1/\ln(\Theta/\Gamma)$.
QLB(2) denotes the two-moment approximation ($L=2$) which is identical with the quantum Lenard-Balescu equation.
The virial expansions for the Ziman formula and for the Lorentz model plasma (2 moment approximation) are also shown.
\label{fig:virial1}}
\end{figure}

These virial plots were created for various examples, see Fig. \ref{fig:virial1}.
The high-temperature limit (\ref{benchsigma}) is shown \cite{Karachtanov16}.
In addition, results from the quantum Lenard-Balescu equation are also shown.
Also shown are the Ziman formula and the Lorentz model plasma, see \cite{cond25}.

It is interesting to compare this with simulations for calculating the Kubo formula (\ref{Kubo}).
The most advanced simulations to date are DFT-MD simulations \cite{Desjarlais02,Mazevet05,Holst11,French2017,Desjarlais,Gajdos2006},
 which were performed at extremely high temperatures and low densities in order to obtain low values for the parameter $x_{\rm Born}$ \cite{RSRB21,French22}.
Without going into the details of these simulations, we point out that the virial plot shows that the contribution of electron-electron collisions is not correctly taken into account.

We summarize the results of the DFT-MD simulations:
\begin{enumerate}
\item
The simulations are highly accurate and show a strictly linear relationship when the particle numbers are sufficiently high.
\item 
The extrapolation $\lim x_{\rm Born} \to 0$ does not yield the Spitzer value, but rather the Lorentz value. This means that the contribution of the $e - e$ collisions is missing.
\item
The slopes of the DFT-MD simulations deviate from the value 0.4917.
 This suggests that dynamical screening is taken into account in different ways.
 \item
 It would be very important to perform PIMC simulations for the DC conductivity of hydrogen plasma.
 \GR{ Similar to the time-dependent density-density correlation function, which indicates the dynamic structure factor, the time-dependent current-current correlation function provides the conductivity according to the Kubo formula (\ref{Kubo}).}
 This could solve the problem of the exact  behavior of conductivity at low density, in particular the determination of the second virial coefficient $\rho_2(T)$. 
 \end{enumerate}
 
There are various approaches to improving DFT-MD simulations for warm and dense matter.
Time-dependent DFT has been applied, see \cite{Cangi24} and references given there.
A more accurate treatment of electron-electron interactions by including the exchange correlation kernel is intended, but this approach does not allow for electron-electron collisions, which lead to the Spitzer limit for low densities.
 
 In order to take into account the contribution of $e - e$ collisions to the conductivity of hydrogen plasmas, a correction factor was introduced in \cite{Heidi,R23,pop24,cond25,Starrett17,Starrett20,Starrett20a} when performing DFT-MD calculations.
The change in the first virial coefficient $\rho_1(T)$ is easy to implement. The correction of $\rho_2(T)$ is quite complicated \cite{cond25}. 
As with the thermodynamic properties of H plasmas, a chemical model based on the Saha equation would be useful to cover a large range of $T,n$ in the calculation of conductivity.

\subsection{Dielectric function}
\label{sec:df}

A more difficult (and more general) response quantity is the dielectric function
\begin{eqnarray}
\label{epspi}
\epsilon(q, \omega)=1 +\frac{i}{\epsilon_0\,\omega}\sigma(q, \omega) = 1 - \frac{e^2}{\epsilon_0 q^2 }\Pi(q, \omega).
\end{eqnarray}
The dielectric function of plasmas contains a lot of informations, including thermodynamic, transport, and optical properties \cite{KKER86}.
The dc conductivity $\sigma_{\rm dc}=\lim_{\omega \to 0} \sigma(0,\omega)$ is a limiting case of the dynamical, wave-vector depending conductivity $\sigma(q, \omega)$. 

%The polarization function $\Pi(q, \omega)$ is related to the dynamic structure factor.
%A well-known result is the RPA expression for the dielectric function, but virial expansions are not yet known, and the analytical behavior near $q,\omega \to 0$ is complex.

The dielectric function $\epsilon(q,\omega)$ is related to the 
%dynamical, wave-vector depending conductivity $\sigma(q, \omega)$  and 
non-local collision frequency according
\begin{equation}
\epsilon(q,\omega)=1-\frac{\omega_{\rm pl}^2}{\omega^2+i\omega\,\nu(q,\omega)},
\end{equation}
where 
\begin{equation}
\omega_{\rm pl}=\left(\frac{e^2 n}{\epsilon_0 m}\right)^{1/2}
\end{equation}
denotes the plasma frequency, $n$ the (free) electron density which equals the proton density for the charged neutral Hydrogen plasma, 
$m$ is the reduced mass. 

The polarization function $\Pi(q, \omega)$  is the irreducible part of the density-density correlation function $\chi(q, \omega)$, which gives after Fourier transformation in space and time the dynamical structure factor \cite{RW1998} (the charge density is decomposed in the densities of the constituents, for simplicity we consider in the following only one component (UEG) and  drop the species)
\begin{eqnarray}
\label{DSF}
S(q,\omega)&=&\frac{1}{2 \pi N} \int_{-\infty}^\infty dt \langle \hat \rho_{\bf q} \hat \rho_{-\bf q}(t)\rangle e^{i \omega t}
\nonumber \\ &&=-\hbar \frac{{\rm Im}\chi (q,\omega-i \eta)}{\pi n(1-e^{-\beta \hbar \omega})}\,.
\end{eqnarray}
Here, $\hat \rho_{\bf q}=\sum_{\bf k}a^+_{{\bf k}-{\bf q}/2}a^{}_{{\bf k}+{\bf q}/2}$ 
are the Fourier components of the electron density, the time dependence is according to the Heisenberg picture. 
As well known from thermodynamic Green function theory \cite{KKER86}, 
$(1/n){\rm Im}\chi (q,\omega-i \eta)$ is the spectral function of the density-density (particle-hole) Green function which can be expressed by Feynman diagrams. 
%As for the single particle propagator where the self-energy (mass operator) is introduced via a Dyson equation, 
A Dyson equation can be given for the  density-density correlation function $\chi(q,\omega)$ in frequency space,
\begin{equation}
\chi (q,\omega)=\frac{\Pi (q,\omega)}{1-\frac{e^2}{\epsilon_0 q^2}\Pi (q,\omega)}
\end{equation}
which shows that the polarization function is like the mass operator for the Coulomb propagator.
The retarded polarization function $\Pi(q,\omega)$ is related to the density response function as
\begin{equation}
\Pi(q,\omega)=\epsilon(q,\omega) \chi(q,\omega)=\frac{\chi(q,\omega)}{1+\frac{ e^2}{\epsilon_0q^2}\chi(q,\omega)}
\end{equation}
so that with relation (\ref{epspi}) 
the dielectric function can be deduced from the density-density correlation function $\chi (q,\omega)$.

Relation (\ref{epspi}) can be rewritten as
\begin{equation}
\epsilon^{-1}(q,\omega)=1+\frac{e^2}{\epsilon_0 q^2} \chi(q,\omega),
\end{equation}
so that from (\ref{DSF}) follows
\begin{equation}
S(q,\omega)
=-\frac{\epsilon_0\hbar q^2}{\pi e^2 n}\frac{{\rm Im}\left[\epsilon^{-1}(q,\omega-i \eta)\right]}{1- e^{-\beta \hbar \omega}},
\end{equation}
%[DraftMS23 (3)]
the fluctuation-dissipation theorem which relates equilibrium density fluctuations to the dielectric function.\\

Within the many-body approach, a main problem is the evaluation of the polarization function $\Pi(q,\omega)$. 
We can start from an approximation $\Pi_0(q,\omega)$ and put all corrections as 
\begin{equation}
\Pi(q,\omega)=\frac{\Pi_0(q,\omega)}{1+\frac{e^2}{\epsilon_0q^2} G_0(q,\omega)\Pi_0(q,\omega)}.
\end{equation}
We have defined a dynamic local field correction $G_0(q,\omega)$ that depends on the choice of $\Pi_0(q,\omega)$.

A standard expression for $\Pi_0(q,\omega)$ is the random-phase approximation $\Pi^{\rm RPA}(q,\omega)$ 
%(Lindhard function) which is the solution in lowest order of interaction,
\begin{equation}
\label{PiRPA}
\Pi^{\rm RPA}(q,\omega)=2\int \frac{d^3k}{(2 \pi)^3}\frac{f(E_{{\bf k-q}/2})-f(E_{{\bf k+q}/2})}{\hbar(\omega+i \eta)+E_{{\bf k-q}/2}-E_{{\bf k+q}/2}}
\end{equation}
as the polarization function of the collision-free plasma (Lindhard function). The limit $\eta \to 0$ is taken to define the contribution near the poles. 
$E_k=\hbar^2k^2/2m$ denotes the kinetic energy of the electron. 
The Fermi distribution functions $f(E)$ can be approximated in the low-density, high-temperature range considered here by the 
classical limit $(n/2) (2 \pi \hbar^2/ m k_{\rm B}T)^{3/2}\exp[-E/k_{\rm B} T]$. We consider here only the contribution of the electrons, the ions contribute only in the low-frequency range, see \cite{R1998}. The factor 2 is owing to spin.

The random phase approximation represents the lowest order of interaction, all higher order contributions are collected in the dynamical local field correction $G^{\rm RPA}(q,\omega)$. 
%It is related to the XC kernel as 
%$K_{\rm XC}(k,\omega)=-G^{\rm RPA}(k,\omega)/(\epsilon_0k^2)$.
For known $\epsilon(q,\omega)$ and $\Pi^{\rm RPA}(q,\omega)$, the dynamical local field correction $G^{\rm RPA}(q,\omega)$ is determined by the relation
\begin{equation}
\epsilon(q,\omega)=1-\frac{e^2}{\epsilon_0 q^2}\frac{\Pi^{\rm RPA}(q,\omega)}{1+\frac{e^2}{\epsilon_0q^2} G^{\rm RPA}(q,\omega)\Pi^{\rm RPA}(q,\omega)}.
\end{equation}
For known $\chi(q,\omega)$ and $\Pi^{\rm RPA}(q,\omega)$, the dynamical local field correction $G^{\rm RPA}(q,\omega)$ is determined by the relation
\begin{equation}
\chi(q,\omega)=\frac{\Pi^{\rm RPA}(q,\omega)}{1-\frac{e^2}{\epsilon_0q^2} [1-G^{\rm RPA}(q,\omega)]\Pi^{\rm RPA}(q,\omega)}.
\end{equation}
Note that other choices of $\Pi_0(k,\omega)$ such as the Kohn-Sham approximation $\Pi^{\rm KS}(q,\omega)$ 
discussed in context with the dynamical conductivity \cite{French22} will also lead to another expressions $G^{\rm KS}(q,\omega)$ for the dynamical local field correction.

The introduction of different quantities is related to the fact that different approximations can be performed.
Similar to the self-energy for the single-particle propagator, partial summations and perturbation expansions are possible.
% Also $\Pi$ and $\chi$ or $S$ are connected like a mass operator. Different limits. The analytical behavior may become more simple, if the perturbation expansion near the ideal, non-interacting limit is not analytic in powers of perturbation.
%We have
%\begin{equation}
%\sigma(k,\omega)=i  \epsilon_0\omega\frac{\frac{e^2}{\epsilon_0 k^2}\Pi^{\rm RPA}(k,\omega)}{1+\frac{e^2}{\epsilon_0k^2} G^{\rm RPA}(k,\omega)\Pi^{\rm RPA}(k,\omega)}
%=i e^2 \frac{ \omega}{k^2} \Pi(k,\omega).
%\end{equation}
The collision frequency is related to the response function according
\begin{eqnarray}
\label{nuG}
\nu(q,\omega)&=&-i \frac{n}{m}\frac{q^2}{\omega}\frac{1}{\chi(q,\omega)}-i\frac{ne^2}{\epsilon_0 m \omega}+i \omega \\ &&
=-i \frac{n}{m}\frac{q^2}{\omega}\frac{1}{\Pi(q,\omega)}+i \omega\nonumber \\
&=&-i\frac{e^2n}{\epsilon_0m \omega}G^{\rm RPA}(q,\omega)-i\frac{n q^2}{m \omega} \frac{1}{\Pi^{\rm RPA}(q,\omega)}+i \omega.\nonumber
\end{eqnarray}
This quantity allows to perform a perturbation expansion.
%These relations allow to calculate the collision frequency if $\chi(k,\omega)$ or $\Pi(k,\omega)$ is known.\\

\subsection{Virial expansion for the dielectric function}

In the low-density limit, at fixed temperature, the plasma behaves like an ideal classical gas, interaction effects become weak.
The polarization function of the ideal fermion gas (only electrons) is given by the RPA expression (\ref{PiRPA}) which reads in the classical limit
\begin{equation}
\Pi^{\rm RPA}(q,\omega)=-\frac{n}{k_{\rm B}T}[1+zD(z)],
\end{equation}
with
\begin{equation}
 z=\frac{\omega}{q}\sqrt{\frac{m_e}{2k_{\rm B}T}}
\end{equation} 
and 
\begin{equation}
\label{D(z)}
D(z)=\frac{1}{\sqrt{\pi}}\int_{-\infty}^\infty dx \frac{e^{-x^2}}{x-z-i \eta}=i \sqrt{\pi}e^{-z^2}[1+{\rm erf}(iz)].
\end{equation}
For the ions we have an additional contribution to $\Pi^{\rm RPA}(q,\omega)$ replacing $m_e$ by $m_i$ in $z$, and for a degenerate electron gas we have to introduce Fermi distribution functions in the Dawson integral (\ref{D(z)}), see Ref. \citep{R1998}.
\GR{ We consider the RPA result to be exactly in the low-density limit case, which is known in analytical form.
It should serve as the lowest-order term in a virial expansion.
Note that an analytical expression can also be given for the degenerate case \cite{Arista84}.}

To find a virial expansion for $\epsilon(q,\omega)$ or the equivalent quantities $\Pi(q,\omega), \sigma(q,\omega)$ is not appropriate 
as shown above for $\sigma_{\rm dc}$, where the virial expansion was considered for the inverse dc conductivity $1/\sigma_{\rm dc}=\rho_{\rm dc}$.
In analogy we consider the virial expansion for the inverse quantities. In particular, we consider the virial expansion for $1/\Pi(q,\omega;T,n)$.

A virial expansion means that we expand in powers of density, so that the virial expansion of  $1/\Pi(q,\omega;T,n)$ includes all powers of density $n$.
As known from Coulomb systems, because of the long-range character of the Coulomb interaction, we have also other dependences on density, 
in particular $n^{1/2}$ because of screening, and $\ln(1/n)$ as shown above for the Coulomb logarithm for the collision integral.
We consider the virial expansion for $1/\Pi(q,\omega)$ in the form
\begin{eqnarray}
\label{vir/Pi}
&&\frac{1}{\Pi(q,\omega)}= \pi_0(q,\omega;T) \frac{1}{n}+\pi_D(q,\omega;T)\frac{1}{n^{1/2}}\nonumber \\
&&+\pi_1(q,\omega;T)\Lambda(T,n) +\pi_2(q,\omega;T)+{\cal O}(n^{1/2}\Lambda).
\end{eqnarray}
In this work, we generalize the virial expansion replacing the term $\ln(1/n)$ by the Coulomb logarithm $\Lambda$,
\begin{eqnarray}
\label{Lamb}
\Lambda(T,n)&=&\ln(1+b)-\frac{b}{1+b},\nonumber \\
 b(T,n)&=&\frac{3 (k_{\rm B}T)^2}{\pi n} \, \frac{4 \pi \epsilon_0 m}{e^2 \hbar^2}.%=\frac{n_\Lambda(T)}{n} .
\end{eqnarray}
This quantity comes from kinetic theory solving the Boltzmann equation for the hydrogen plasma in Born approximation. 

The lowest order virial coefficient
\begin{equation}
\pi_0(q,\omega;T)=\frac{n}{\Pi^{\rm RPA}(q,\omega;T)}=-\frac{k_{\rm B}T}{1+zD(z)}
\end{equation}  
is exactly known.  The virial coefficient $\pi_D(q,\omega;T)$ contains the Debye shifts of the quasi-particle energies. If the Debye shifts do not depend on the momentum $\hbar {\bf p}$  in Eq. (\ref{PiRPA}), they can be absorbed by the chemical potential,  and we omit this contribution.

To extract the virial coefficient $\pi_1(q,\omega;T)$, one could create a virial plot for 
\begin{equation}
\pi^{\rm eff}_1(q,\omega;T,n)=\left[\frac{1}{\Pi(q,\omega)}- \frac{1}{\Pi^{\rm RPA}(q,\omega)}\right] \frac{1}{\Lambda(T,n)}
\end{equation}
as function of $x=1/\Lambda $. 
Because of the virial expansion (\ref{vir/Pi}),
\begin{equation}
\pi^{\rm eff}_1(q,\omega;T,n)=\pi_1(q,\omega;T)+\pi_2(q,\omega;T) \frac{1}{\Lambda(T,n)}+{\cal O}(n^{1/2})
\end{equation}
the virial plot allows to determine the virial coefficients $ \pi_1(q,\omega;T), \pi_2(q,\omega;T)$ from the value and the slope of $\pi^{\rm eff}_1(q,\omega;T,n)$ at $x=0$.
The elaboration of the viral expansion of the polarization function is very complex and goes beyond the scope of this work.

Note that 
\begin{equation}
\frac{1}{\Pi(q,\omega)}- \frac{1}{\Pi^{\rm RPA}(q,\omega)}=\frac{e^2}{\epsilon_0 q^2} G^{\rm RPA}(q, \omega)
\end{equation}
so that we are discussing a virial expansion of the dynamical local field corrections or the collision frequency, eq. (\ref{nuG}):
\begin{eqnarray}
\label{nuPi}
&&\nu(q,\omega;T,n)=-i\frac{q^2 n}{m_e \omega} \, \frac{1}{\Pi(q,\omega)}+i\omega \nonumber \\
&&=\nu_0(q,\omega; T)+\nu_D(q,\omega; T)n^{1/2}+\nu_1(q,\omega; T)n \Lambda(T,n) \nonumber \\
&& +\nu_2(q,\omega; T)n+{\cal O}(n^{3/2}\Lambda(T,n)).
\end{eqnarray}
%To extract $\nu_1(k,\omega; T),\nu_2(k,\omega; T)$, we plot $[-i\frac{k^2 n}{m_e \omega} \, \frac{1}{\Pi(k,\omega)}+i\omega-\nu_0(k,\omega; T)-\nu_D(k,\omega; T)n^{1/2}]/ \Lambda(T,n)$ 
%as function of $x=1/\Lambda $, as above.
The virial coefficients for $1/\Pi(q,\omega;T,n), G(q, \omega;T,n)$ or $\nu(q,\omega;T,n)$ are functions of $q,\omega$ in addition to $T$.
%In the following, we present virial plots for analytical calculations as well as for the results of numerical simulations.

\subsection{Analytical expressions for the plasma response}

Analytical expressions are obtained from generalized linear response theory.
We introduce moments of the distribution function \cite{R1998,RW1998}
\begin{equation}
\label{Bc}
B^c_{n,q}=\sum_p\left(\frac{\hbar^2}{2 m_c k_{\rm B}T}\right)^{(2n+1)/2} p^{2n}p_z \hat n^c_{p,q}\end{equation}
with $\hat n^c_{p,q}=a^+_{c,p-q/2}a_{c,p+q/2}$, and $n=0,1,2\dots$.
\GR{ The wave vector $\bf p$ has the absolute value $p$ and the $z$-component $p_z$.}
This set of observables includes also the current density operator
\begin{equation}
J_q=\frac{1}{\Omega_0}\sum_{c,p}\frac{e_c}{m_c} \hbar p_z \hat n^c_{p,q}.
\end{equation}
As known from the Chapman-Enskog approach to solve the kinetic equations, increase the number of moments will improve the result.
For $q=0$, this approach to the dynamical conductivity has been studied in detail, see \cite{Reinholz12}.

For the Hydrogen plasma, the calculation of the dielectric function for the fully ionized Hydrogen plasma 
has been given in the one-moment approximation for the distribution of electrons and ions in \cite{R1998}, 
and the two-moment approximation in \cite{RW1998}.

For simplicity, we consider the Born-Oppenheimer approximation where ions at fixed configuration are treated as external field. 
We drop $c$ considering only electrons (spin is not treated explicitly).\\

Within a generalized linear response approach, the polarization function is found as \cite{RW1998}
\begin{eqnarray}
\label{PiM1}
&&\Pi(q,\omega)=i\frac{q^2}{\omega}\epsilon(q,\omega)
\begin{vmatrix}
    0 & M_{0n}(q,\omega) \\ 
     M_{m0}(q,\omega) &M_{mn}(q,\omega) 
\end{vmatrix}/\begin{vmatrix}M_{mn}(q,\omega) 
\end{vmatrix}\nonumber \\
&&{}
\end{eqnarray}
with (the indices $c$ and $q$ in Eq. (\ref{Bc}) are omitted)
\begin{equation}
M_{0n}(q,\omega)=(J_q|B_n),\qquad M_{m0}(q,\omega)=(B_m|J_q)
\end{equation}
and
\begin{eqnarray}
\label{PiM2}
&&M_{mn}(q,\omega) = -i \omega (B_m|B_n)-(\dot B_m|B_n)+
\langle \dot{B}_m;\dot B_n \rangle_{\omega+i \eta}\nonumber \\
&&+ 
\begin{vmatrix}
    0 & \langle \dot B_m; B_j \rangle_{\omega+i \eta} \\ 
      \langle B_i; \dot B_n \rangle_{\omega+i \eta} &\langle B_i; B_j \rangle_{\omega+i \eta} 
\end{vmatrix}/\begin{vmatrix}\langle B_i; B_j \rangle_{\omega+i \eta}
\end{vmatrix}.
\end{eqnarray}
The equilibrium correlation functions are defined as
\begin{eqnarray}
(A|B)&=&\frac{1}{\beta}\int_0^\beta d\tau {\rm Tr}\left[\rho_{\rm eq} A(-i\hbar \tau)B^+\right], \nonumber\\
\langle A; B \rangle_z&=&\int_0^\infty e^{izt} (A(t)|B).
\end{eqnarray}
The equilibrium correlation functions ($\rho_{\rm eq}$ is the statistical operator, Eq. (\ref{rhoeq}))
can be calculated within perturbation theory using the method of thermodynamic Green functions.
Numerical simulations can be used as an alternative to avoid perturbation  expansions: DFT-MD simulations,
 which approximate electron-electron correlations, and exact PIMC simulations, which are, however,
very computationally expensive.

In Refs. \cite{R1998,RW1998} several approximations were made in order to obtain analytical expressions for the dynamical collision frequency: 
a finite number of moments was considered
(rapid convergence with increasing number of moments was shown for the dc conductivity \cite{Redmer97}).
The perturbation expansion of the correlation functions was limited to the lowest orders of the interaction potential, neglecting self-energy and related effects.
A classical limit was assumed, replacing Fermi distributions with Boltzmann distributions, which is possible at low densities 
and can be improved immediately with little numerical effort, but only affects higher-order virial coefficients.

\subsection{Relation to thermodynamics}
To evaluate thermodynamic properties of plasmas, we investigate the isothermal compressibility $\kappa_{\rm iso}$,\begin{equation}
\kappa_{\rm iso}(T,\mu) = -\frac{1}{\Omega_0}\, \frac{\partial \Omega_0}{\partial p}|_{T,N}=\frac{1}{n^2} \frac{\partial n}{\partial \mu}|_T.
\end{equation}
For simplicity, we will only consider one-component systems, where $\Omega_0$ denotes the volume. If $\kappa_{\rm iso}$ is known, integration yields $n(T,\mu)$,
and further integration yields the pressure,
\begin{equation}
p(T,\mu) =\int_{-\infty}^\mu n(T,\mu') d\mu'\,.
\end{equation}
We have $p(T,\mu) \Omega_0$ as thermodynamic potential. All other thermodynamic variables can be derived from the thermodynamic potential.

The  isothermal compressibility $\kappa_{\rm iso}$ is related to the response function $\chi(q,\omega)$ as
\begin{equation}
\kappa_{\rm iso}(T,\mu)=\frac{1}{n^2T} \lim_{q \to 0} \int_{-\infty}^\infty \frac{d \omega}{\pi} \frac{1}{e^{\omega/T}-1} {\rm Im} \chi(q,\omega)\,.
\end{equation}
This can also be expressed by the static structure factor,
\begin{equation}
S({ q})=\int \frac{d \omega}{2 \pi} S({ q}, \omega),
\end{equation}
and
\begin{equation}
S({ q}\to 0)=(\langle \hat n^2\rangle-n^2)/n=T\frac{\partial n}{\partial \mu}|_T.
\end{equation}

The virial expansion is also possible for these quantities. 
Correlations, in particular bound states, are implemented in the polarization function using a cluster decomposition of the polarization function $\Pi(q,\omega)$ \cite{RD79}. The connection to the Beth-Uhlenbeck formula is given in Ref. \cite{Landau18}.

\subsection{Numerical simulations}
PIMC calculations were performed for the imaginary-time density-density correlation function
\begin{equation}
F_{ij}(q, \tau ) = \langle \hat n_i(q, \tau ) \hat n_j(-q, 0)\rangle,
\end {equation}
see Ref. \cite{Groth19,Dornh24,Bellenbaum2025} and other references cited therein.
However, it has been difficult to obtain results for the entire $\{q,\omega\}$ plane.

A fundamental quantity for partially ionized plasmas, which is not precisely defined, is the density of free electrons $n^*_e$ or the degree of ionization $n^*_e/n_e$. 
If there are precisely defined bound states and scattering states for the electrons,
it seems possible to divide the total number of electrons into localized and itinerant electrons.
However, these concepts become invalid in dense systems, where we have a finite lifetime of bound states and resonances in the continuum.
It would be interesting to relate the free electron density to an observable quantity of the plasma. 
In \cite{RCER25}, it was proposed to use the long-wavelength limit.
\begin{equation}
\label{polfct}
\Pi(q, \omega=0) = \epsilon_r \epsilon_0 e^{-2} \kappa_{\rm free}^2+(\epsilon_r-1)\epsilon_0e^{-2} q^2+ {\cal O}(q^4).      
\end{equation}
The first term with $\kappa_{\rm free}$ is determined by the free quasiparticle density, whereas the second term contains via $\epsilon_r$ the bound states (local field correction, Clausius-Mossotti term).
In the low-density, nondegenerate limit considered here, we have with $\epsilon_r \approx 1$
\begin{equation}
    \kappa^2_{\rm free}=\frac{2 e^2 n^*}{\epsilon_0 k_{\rm B}T}.
\end{equation}
In this way, $S(q)$, which is accessible from PIMC simulations, can define the free electron density, which is relevant for screening but also for other properties.

A more detailed investigation of the dynamic structure factor and related quantities such as the dielectric function is an important contribution to the understanding of warm dense plasmas. There are numerous contributions from both quantum statistics and numerical methods. A comparison of both approaches and the derivation of exact relations, such as virial expansions, is an interesting future project. 

\subsection{Open problems}
\begin{enumerate}
\item
Current calculations of WDM conductivity are performed using DFT-MD simulations.
However, $e-e$ collisions, which are important in the low-density range, are not taken into account.
PIMC simulations are urgently needed to obtain values for conductivity in the low-density range.

\item
The second virial coefficient $\rho_2(T)$ is of interest which is determined by screening and strong collisions. Its dependence on $T$ is currently unknown. PIMC simulations can be used to investigate the dependence on $T$.

\item
As the temperature $T$ decreases, bound states are formed that influence conductivity. 
An open question is how the second virial coefficient $\rho_2(T)$ is influenced by the formation of bound states.

\item
Of interest are the higher orders of virial expansion. \GR{In addition to the Debye-Onsager relaxation effect, higher-order terms describing the effects of degeneracy are also of interest.} They enable comparison with experimental data.

\item
The dielectric function and the associated dynamic structural factors contain a wealth of information about the properties of plasmas. Virial expansions can be derived to obtain benchmarks in limiting cases.

\end{enumerate}

\section{Conclusions}
\label{sec:concl}

Benchmarks from analytical approaches are important for checking the quality of numerical data.
As examples, we have considered the equations of state and the conductivity of H plasmas.
Numerical simulations, on the other hand, can help to verify analytical results, such as the correct form of the $D$ function for the second virial coefficient and the question of whether $e-e$ collisions are taken into account in DFT-MD simulations of plasma conductivity.
High-accuracy numerical simulations can be used to extract further higher-order coefficients of the virial expansion.
They also show the validity range of the virial expansions.
Both together are relevant for the formulation of interpolation formulas that satisfy the exact low-density limits.

\GR{
In this article, which is partly a review paper, we demonstrate a general method of virial expansion for various physical properties, in the expectation that there will be a unified theory in the future.
Furthermore, we can identify which parameter values are of interest for new, accurate simulations in order to gain better insights, for example to obtain higher-order virial coefficients or to address difficult problems such as the description of bound state formation, the degree of ionisation and the ionization potential depression.
We have presented new results on the convergence of virial expansion for the UEG.
New results on the thermodynamics and transport coefficients of H plasmas are expected as soon as new high-precision simulations become available.
The virial expansion of the dielectric function is highlighted as a challenging project for the future.
Systematic virial expansions in combination with numerical simulations can give us a better understanding of the properties of dense plasmas and warm dense matter.}

\appendix

\section{Data for the UEG}
\label{app:UEG}
We collect data for the UEG published in \cite{TD} and \cite{R24} and used for Figures \ref{fig:1} to \ref{fig:v2} in Section \ref{sec:UEG}.
The PIMC calculations were performed by the group led by T. Dornheim and J. Vorberger.

\begin{widetext}

\begin{table}[h]
\begin{center}
\hspace{0.5cm}
 \begin{tabular}{|c|c|c|c|c|c|c|}
\hline
\# &$r_s$& $n_{\rm Bohr}$ & $\Theta$ & $T$ [K]& $T_{\rm Ha}$ & $v^{\rm PIMC}$ [Ha]  \\ 
\hline
1& 0.5    &1.9099 &    128.0 &2.977$\times 10^8$ &  942.9 	&- 0.082621425 $\pm$ 0.00012815\\
2&& &64.0			&1.488$\times 10^8$ & 471.4 & - 0.11804562 $\pm$  0.0001239\\
3&& &32.0			& 7.444$\times 10^7$ & 235.7 & - 0.16927203 $\pm$  5.788 e - 05\\
4&& &16.0			& 3.722$\times 10^7$ & 117.8 & - 0.24239928 $\pm$  0.0001579\\
5&& &8.0 			& 1.8609$\times 10^7$ & 58.93 &- 0.34476409 $\pm$  0.0004249\\
\hline
6&2 & 0.02984 &128.0			& 1.861$\times 10^7$ &58.93&  - 0.04022477 $\pm$  4.794 e - 05\\
7&& &64.0 			& 9.304$\times 10^6$ & 29.47 &- 0.05680616 $\pm$ 2.679 e - 05\\
8&& &32.0 			&4.652$\times 10^6$  & 14.73 &- 0.07971471 $\pm$  2.137 e - 05\\
9&& &16.0 			&2.326$\times 10^6$  & 7.366 &- 0.11012571 $\pm$  7.846 e - 05\\
10&& &8.0				& 1.1630$\times 10^6$ & 3.683 & - 0.14866113 $\pm$ 0.0001037\\
\hline
11&20& 2.984$\times 10^{-5}$ & 128.0 		& 186090& 0.5893 &- 0.01192989 $\pm$  9.436e - 06\\
12&& &64.0 			&93044.8 & 0.2947 &- 0.01600505 $\pm$  3.663 e - 06\\
13&& &32.0 			&46522.4 & 0.1473 &- 0.02071120 $\pm$ 2.067 e - 06\\
14&& &16.0			&23261.2 & 0.07366&- 0.02563372 $\pm$ 3.281 e - 06\\
15&& &8.0				& 11630.6& 0.03683&- 0.03020982 $\pm$ 2.283 e - 06\\
\hline
16&40&3.73$\times 10^{-6}$ &    512     & 186090.& 0.5893 &-0.00434040 $\pm$  0.00000860      \\
17&40 &3.73$\times 10^{-6}$ & 256          & 93044.8& 0.2947 & -0.00597188 $\pm$  0.00001116 \\
18&12.5 &1.222$\times 10^{-4}$ &	50	&186090. & 0.5893 & -0.02338182 $\pm$  0.00000916 \\
19&12.5 &1.222$\times 10^{-4}$ & 	25	&93044.8 & 0.2947 & -0.03060253 $\pm$  0.00000913  \\
20&2 &0.02984 & 217.2     & 3.158$\times 10^7$ & 100 & -0.03085446  $\pm$  0.00006612  \\
21&1 &0.2387 &	54.3    &3.158$\times 10^7$  & 100 &-0.08899252  $\pm$  0.00027539  \\
\hline
 \end{tabular}
\caption{PIMC calculations for the UEG: $v^{\rm PIMC}(T,n)$ . $n_{\rm Bohr}=3/(4 \pi r_s)^3$, $T_{\rm Ha}=(9 \pi/4)^{2/3} \Theta /(2 r_s^2)$, $T=T_{\rm Ha}\times 315777.1 {\rm K}$.}
\label{Tab:4}
\end{center}
\end{table}

\end{widetext}

{\bf Acknowledgments}

The author. thanks M. Bethkenhagen, M. Bonitz, T. Dornheim, W. Ebeling, W.D. Kraeft,  R. Redmer, H. Reinholz, J. Vorberger for discussions.
%%%%%%%%%%%%%%%%%%%%%%%%%%%

%%%%%%%%%%%%%%%%%%%

\begin{thebibliography}{10}
%\footnotesize
\renewcommand{\baselinestretch}{0.5}


\bibitem{Bonitz24}
  M. Bonitz, J. Vorberger, M. Bethkenhagen, M. Böhme, D. Ceperley, A. Filinov, T. Gawne, F. Graziani, G. Gregori, P. Hamann, S. Hansen, M. Holzmann, S. X. Hu, H. Kählert, V. Karasiev, U. Kleinschmidt, L. Kordts, C. Makait, B. Militzer, Z. Moldabekov et al., {\it Toward first principles simulations of dense hydrogen}, Phys. Plasmas {\bf 31}, 110501 (2024).
  
  \bibitem{WDM}
J.  Vorberger, F. Graziani, D. Riley, A. D. Baczewski, I. Baraffe, M. Bethkenhagen, S. Blouin, M. P. Böhme, M. Bonitz, M. Bussmann, et al.,
   {\it Roadmap for warm dense matter physics}, arxiv:2505.02494 (2025).
  
  \bibitem{B20}
  M. Bonitz, T. Dornheim, Zh.A. Moldabekov, S. Zhang, P. Hamann, H. Kählert, A. Filinov, K. Ramakrishna, J. Vorberger,
 {\it Ab initio simulation of warm dense matter}, Phys. Plasmas {\bf 27}, 042710 (2020).
  
  \bibitem{FW}
A. L. Fetter and J. D. Walecka, {\it Quantum Theory of Many-Particle Systems} (McGraw-Hill, 1971).

\bibitem{Buch}
G. R\"opke,  {\it Nonequilibrium Statistical Physics} (Wiley-VCH Verlag GmbH, 2013).

 \bibitem{KB}
Kadanoff and G. Baym, {\it Quantum Statistical Mechanics:
Green's Function Methods in Equilibrium and Nonequilibrium Problems} (W.A. Benjamin, New York  1962).

\bibitem{KKER86}
W.-D. Kraeft, D. Kremp, W. Ebeling, and G. R\"opke, 
{\it Quantum Statistics of Charged Particle Systems} 
(Plenum Press, New York and London, 1986).

\bibitem{cond25}
G. R\"opke,  R. Redmer, M. Sch\"orner, H. Reinholz, U. Kleinschmidt, and M. Bethkenhagen,
{\it Electron-electron collisions in calculations of the electrical conductivity for warm dense matter based on density functional theory},
Phys. Rev. E {\bf 111}, 055201 (2025).

\bibitem{RCER25}
G. R\"opke, C.L. Lin, W. Ebeling, H. Reinholz, {\it The virial expansion of the Hydrogen equation of state in comparison to PIMC simulations: The quasiparticle concept, IPD, and ionization degree}, 
Contrib. Plasma Phys. e70078 (published online 2026).
\bibitem{pop24}
G. R\"opke, {\it Electrical conductivity of hydrogen plasmas:
Low-density benchmarks and virial expansion
including e–e collisions},
Phys. Plasmas {\bf 31}, 042301 (2024).

\bibitem{RKKKZ78} 
G. R\"opke, K. Kilimann,
W. D. Kraeft, D. Kremp and R. Zimmermann, {\it The influence of dynamical effects on the two-particle states
(excitons) in the electron-hole plasma}, Phys. Stat. Sol. (b) {\bf 88}, K59 (1978).

\bibitem{ZKKKR78} 
R. Zimmermann, K. Kilimann,
W. D. Kraeft, D. Kremp and G. R\"opke, {\it namical screening and self-energy of excitons in the electron-
hole plasma}, Phys. Stat. Sol. (b) {\bf 90}, 175 (1978).

\bibitem{BU}
E. Beth and G. Uhlenbeck, {\it The quantum theory of the non-ideal gas. II. Behaviour at low temperatures}, Physica {\bf 4}, 915 (1937).

\bibitem{Schmidt90} 
M. Schmidt, G. R\"opke, and H. Schulz, {\it Generalized Beth-Uhlenbeck approach for hot nuclear matter}, Annals Phys. {\bf 202}, 57
(1990).

\bibitem{Ebeling67}
W. Ebeling, {\it Statistical thermodynamics of bound states in plasmas} (German),
Ann. Phys. (7), {\bf 19}, 104  (1967). 

\bibitem{Ebeling68}
W. Ebeling, {\it The exact free energy of low density quantum plasma}, Physica {\bf  40}, 290 (1968). 

\bibitem{Kahlbaum}
T. Kahlbaum, {\it Quantum diffraction term in the free energy for Coulomb plasma}, J. Phys. (France) {\bf 10}, 455 (2000).

\bibitem{AlMaMono}
A. Alastuey, P.A. Martin, {\it Statistical Mechanics of Coulomb Systems}, EPFL Press, Lausanne 2025.

\bibitem{Millitzer}
B. Militzer and E. L. Pollock, {\it Variational density matrix method for warm, condensed matter: Application to dense hydrogen}, Phys. Rev. E {\bf 61}, 3470 (2000).

\bibitem{Dornheim2018}
T. Dornheim, S. Groth, and M. Bonitz, {\it The uniform electron gas at warm dense matter conditions},
Physics Reports {\bf 744}, 1 (2018).

\bibitem{PIMC1}
T. Dornheim and J. Vorberger, {\it Finite-size effects in the reconstruction of dynamic properties from ab initio path integral Monte Carlo simulations},
Phys. Rev. E {\bf 102}, 063301 (2020).

\bibitem{Dorn25}
T. Dornheim, T. Döppner, P. Tolias, M. P. Böhme, L. B. Fletcher, T. Gawne, F. R. Graziani, D. Kraus, M. J. MacDonald, Zh. A. Moldabekov, S. Schwalbe, D. O. Gericke and J. Vorberger, 
{\it Unraveling electronic correlations in warm dense quantum plasmas}, Nature Comm. {\bf 16}, 5103 (2025).

\bibitem{Filinov23}
A. Filinov,  M. Bonitz,  {\it The EOS of partially Hydrogen and deuterium plasma revisited}, Phys. Rev. E {\bf 108}, 055212 (2023).

\bibitem{TD}
T. Dornheim, J. Vorberger, 
Zh. Moldabekov, G. R\"opke, W.-D. Kraeft, {\it he uniform
electron gas at high temperatures: ab initio path integral
monte carlo simulations and analytical theory},
 High Energy Density Physics {\bf 45}, 101015 (2022).

\bibitem{R24}
G. R\"opke, T. Dornheim, J. Vorberger, D. Blaschke, B. Mahato, {\it Virial coefficients of the Uniform Electron Gas from Path Integral Monte Carlo
Simulations}
Phys. Rev. E {\bf 109}, 025202  (2024).

\bibitem{GDSMFB17}
S. Groth, T. Dornheim, T. Sjostrom, F.D. Malone, W.M.C. Foulkes, M. Bonitz, 
{\it Ab initio exchange–correlation free energy of the uniform electron gas at warm dense matter conditions}, 
Phys. Rev. Lett. {\bf 119}, 135001 (2017) .

\bibitem{KKR15}
W.-D. Kraeft, d. Kremp, G. R\"opke, {\it Direct linear term in the equation of state of plasmas}, 
Phys. Rev. E {\bf 91}, 013108 (2015).

\bibitem{Ebel25}
W. Ebeling, {\it Problems of quantum-statistical thermodynamics of plasmas:
High- and low-temperature limits and analyticity}, Contr. Plasma Phys. {\bf 25},
e202400048 (1925).

\bibitem{Dornheim25}
T. Dornheim, Zh. A. Moldabekov, S. Schwalbe, and J. Vorberger, {\it Direct free energy calculation from ab initio path integral Monte Carlo simulations of warm dense matter},
Phys. Rev.  B {\bf 111}, L041114 (2025).

\bibitem{Zubarev}
D.N. Zubarev, V.G. Morosov, and G. Roepke, {\it Statistical mechanics of nonequilibrium processes I/II} (Wiley 1996/7).

\bibitem{Kubo1957}
R. Kubo, {\it Statistical mechanical theory of irreversible processes. 1. General theory and simple applications in magnetic and conduction problems}, Journal of the Physical Society of Japan {\bf 12}, 570 (1957).

\bibitem{Ropke18}
G. R\"opke, {\it Electrical conductivity of charged particle systems and Zubarev's nonequilibrium statistical operator method}, TMF, {\bf 194}, 90  (2018); Theoret. and Math. Phys. {\bf 194}, 74  (2018).

\bibitem{Roep88}
G. R\"opke, {\it Quantum-statistical approach to the electrical conductivity of dense, high-temperature plasmas}, Phys. Rev. A {\bf 38}, 3001 (1988).

\bibitem{RR89}
G. R\"opke and R. Redmer, {\it Electrical conductivity of nondegenerate, fully ionized plasmas}, Phys. Rev. A {\bf 39}, 907 (1989).

\bibitem{Redmer97}
R. Redmer, {\it Physical properties of dense, low-temperature plasmas}, Physics Reports {\bf 282}, 35 (1997).

\bibitem{Reinholz12}
H. Reinholz and G. R\"opke, {\it Dielectric function beyond the random-phase approximation: Kinetic theory versus linear response theory}, Phys. Rev. E {\bf 85}, 036401 (2012).

\bibitem{Reinholz05}
H. Reinholz, {\it Dielectric and optical properties of dense plasmas}, Annales de Physique {\bf 30}, 1 (2005).

\bibitem{Spitzer53}
J.L. Spitzer and R. H\"arm, {\it Transport phenomena in a completely ionized gas}, Phys. Rev. {\bf 89},  977  (1953).

\bibitem{Karachtanov16}
V.S. Karakhtanov, {\it Corrections to the electrical conductivity of the fully  ionized gas}, 
Contrib. Plasma Phys. {\bf 56},  343  (2016).

\bibitem{R23}
G. R\"opke, {\it Thermodynamic and transport properties of plasmas: Low-density benchmarks}, 
Contrib. Plasma Phys. {\bf 63}, e202300002 (2023).

\bibitem{LeeMore84}
Y.T. Lee and R.M. More, {\it An electron conductivity model for dense plasmas},
Phys. Fluids {\bf 27}, 1273 (1984).

\bibitem{Grabowski20}
P. E. Grabowski, S.B. Hansen, M.S. Murillo, L.G. Stanton, F.R. Graziani, A.B. Zylstra, S.D. Baalrud, P. Arnault,  A.D. Baczewski, L.X. Benedict, et al., {\it Review of the first charged-particle transport coefficient comparison workshop}, 
High Energy Dens. Phys. {\bf 37}, 100905 (2020).

\bibitem{Stanek2024}
L. J. Stanek, A. Kononov, S. B. Hansen, B. M. Haines, S. Hu, P. F. Knapp, M. S. Murillo, L. G. Stanton, H. D. Whitley, S. D. Baalrud, L. J. Babati, A. D. Baczewski, M. Bethkenhagen, A. Blanchet, R. C. Clay, K. R. Cochrane, L. A. Collins, A. Dumi, G. Faussurier, M. French, Z. A. Johnson, V. V. Karasiev, S. Kumar, M. K. Lentz, C. A. Melton, K. A. Nichols, G. M. Petrov, V. Recoules, R. Redmer, G. R\"opke, M. Sch\"orner, N. R. Shaffer, V. Sharma, L. G. Silvestri, F. Soubiran, P. Suryanarayana, M. Tacu, J. P. Townsend, and A. J. White,
 {\it Review of the second charged-particle transport coefficient code comparison workshop,} Phys. Plasmas {\bf 31}, 052104 (2024).


\bibitem{Desjarlais02}
M. P. Desjarlais, J. D. Kress, and L. A. Collins, {\it Electrical conductivity for warm, dense aluminum plasmas and liquids},  
Phys. Rev. E {\bf 66}, 025401(R) (2002).

\bibitem{Mazevet05}
S. Mazevet, M. P. Desjarlais, L. A. Collins, J. D. Kress, and N. H. Magee, {\it Simulations of the optical properties of warm dense aluminum}, 
Phys. Rev. E {\bf 71}, 016409 (2005).

\bibitem{Holst11}
B. Holst, M. French, and R. Redmer, {\it Electronic transport coefficients from ab initio simulations and application to dense liquid hydrogen}, Phys. Rev. B {\bf 83}, 235120 (2011).

\bibitem{French2017}
M. French and R. Redmer, {\it Electronic transport in partially ionized water plasmas}, 
Phys. Plasmas {\bf 24}, 092306 (2017).

\bibitem{Desjarlais}
M.P. Desjarlais, C.R. Scullard, L.X. Benedict, H.D. Whitley, and R. Redmer, {\it Density-functional calculations of transport properties in the non-degenerate limit and the role of electron-electron scattering}, 
Phys. Rev. E {\bf 95}, 033203 (2017).

\bibitem{Gajdos2006}
M. Gajdos, K. Hummer, G. Kresse, J. Furthm{\"u}ller, and F. Bechstedt, {\it Linear optical properties in the projector-augmented wave methodology}, 
Phys. Rev. B {\bf 73}, 045112 (2006).

\bibitem{RSRB21}
G. R\"opke, M. Sch\"orner, R. Redmer, and M. Bethkenhagen, {\it Virial expansion of the electrical conductivity of hydrogen plasmas}, 
Phys. Rev. E {\bf 104}, 045204 (2021).

\bibitem{French22}
M. French, G.~R\"opke, M. Sch\"orner, M. Bethkenhagen, M.P. Desjarlais, and R. Redmer, {\it Electronic transport coefficients from density functional theory across the plasma plane},  
Phys. Rev. E {\bf 105}, 065204 (2022).

\bibitem{Cangi24}
K. Ramakrishna, M. Lokamani, and A. Cangi, {\it Electrical conductivity of warm dense hydrogen from ohm’s law and time-dependent density functional theory}, 
Electron. Struct. {\bf 6}, 045008 (2024).

\bibitem{Heidi}
H. Reinholz, G. R\"opke, S. Rosmej, and R. Redmer, {\it Conductivity of warm dense matter including electron-electron collisions},
Phys. Rev. E {\bf 91}, 043105 (2015).
 
\bibitem{Starrett17}
C.E. Starrett, {\it  Potential of mean force for electrical conductivity of dense plasmas}, High Energy Density Phys. {\bf 25}, 8 (2017).

 \bibitem{Starrett20}
N.R. Shaffer and C.E. Starrett, {\it Model of electron transport in dense plasmas spanning temperature regimes}, 
Phys. Rev. E {\bf 101}, 053204 (2020). 

\bibitem{Starrett20a}
N.R. Shaffer and C.E. Starrett, {\it Correlations between conduction electrons in dense plasmas}, 
Phys. Rev. E {\bf 101}, 013208 (2020).
 
\bibitem{RW1998}
G. R\"opke and A. Wierling, {\it Dielectric function of a two-component plasma including collisions}, Phys. Rev. E {\bf 57}, 7075 (1998).

\bibitem{R1998}
G. R\"opke, {\it Dielectric function and electrical dc conductivity of nonideal plasmas}, Phys. Rev. E {\bf 57}, 4673 (1998).

\bibitem{Arista84}
N. R. Arista and W. Brandt, {\it Dielectric response of quantum plasmas in thermal equilibrium},
Phys. Rev. A {\bf 29}, 1471 (1984).

\bibitem{RD79}
G. R\"opke and R. Der, {\it The influence of two-particle states (excitons) on the dielectric function of the electron—hole plasma}, Phys. Stat. Sol. (b) {\bf 92}, 501 (1979).

\bibitem{Landau18}
G. R\"opke, D.N. Voskresensky, I.A. Kryukov, D. Blaschke, {\it  Fermi liquid, clustering, and structure factor in dilute warm nuclear matter}, 
Nucl. Phys. A {\bf 970}, 224 (2018).

\bibitem{Groth19}
S. Groth, T. Dornheim, and J. Vorberger, {\it Ab initio path integral Monte Carlo approach to the static and dynamic density response of the uniform electron gas}, Phys. Rev. B {\bf 99}, 235122 (2019).

\bibitem{Dornh24}
T. Dornheim, S. Schwalbe, P. Tolias, M. P. B\"ohme, Z. A. Moldabekov, and J. Vorberger, {\it Ab initio density response and local field factor of warm dense hydrogen}, Matter Radiat. Extrem. {\bf 9}, 057401 (2024).

\bibitem{Bellenbaum2025}
H.M. Bellenbaum, M. P. Böhme, M. Bonitz, T. Döppner, L. B. Fletcher, T. Gawne, D. Kraus, Zh. A. Moldabekov, S. Schwalbe, J. Vorberger, T. Dornheim, {\it Estimating ionization states and continuum lowering from ab initio path integral Monte Carlo simulations for warm dense hydrogen}, 
Phys. Rev. Research {\bf 7}, 033016 (2025).



\end{thebibliography}
\end{document}